\documentclass[a4paper,fleqn,usenatbib,useAMS]{mnras}

\usepackage{graphicx}	% Including figure files
\usepackage{tabularx}
\usepackage{amsmath}	% Advanced maths commands
\usepackage{amssymb}	% Extra maths symbols
\usepackage{multicol}        % Multi-column entries in tables
\usepackage{multirow}
\usepackage{bm}		% Bold maths symbols, including upright Greek
\usepackage{pdflscape}	% Landscape pages
\usepackage{footmisc}

\def \aj {{\rm {AJ}}}
\def \araa {{\rm {ARA\&A}}}
\def \apj {{\rm {ApJ}}}

\def \apjs {{\rm {ApJS}}}
\def \apjl {{\rm {ApJL}}}

\def \aap {{\rm {A\&A}}}
\def \aapr {{\rm {A\&AR}}}

\def \mnras {{\rm {MNRAS}}}

\def \pasp {{\rm {PASP}}}

\def \sol {{\rm M$_\odot$}}
\def \Lsol {{\rm L$_\odot$}}
\def \solyr {{\rm M$_\odot$\,yr$^{-1}$}}
\def \arcsec {{\rm $^{\prime\prime}$}}
\def \micron{{\rm \,$\mu$m}}
\def \colhtwo {{$N_\mathrm{H_2}$}}
\def \colhtwowarm {{$N_\mathrm{H_2}^{warm}$}}
\def \colhtwocool {{$N_\mathrm{H_2}^{cool}$}}
\def \HII {{\ion{H}{II}}}

\def \dir {} 

\begin{document}

%										Title											%
\title[Star formation in the Galactic Centre]{Star formation rates and efficiencies in the Galactic Centre}

\author[A.T. Barnes et al.]
{A.T. Barnes$^{1}$\thanks{E-mail: a.t.barnes@2014.ljmu.ac.uk}, S.N. Longmore$^{1}$, C. Battersby$^{2}$, J. Bally$^{3}$, J.M.D. Kruijssen$^{4,5}$,  \and J.D. Henshaw$^{1}$ and D.L. Walker$^{1}$\\
 $^{1}$Astrophysics Research Institute, Liverpool John Moores University, 146 Brownlow Hill, Liverpool L3 5RF, UK\\
 $^{2}$Harvard-Smithsonian Center for Astrophysics, Cambridge, MA, 02138\\
 $^{3}$Centre for Astrophysics and Space Astronomy, University of Colorado, UCB 389, Boulder, CO 8030\\ 
 $^{4}$Astronomisches Rechen-Institut, Zentrum f\"{u}r Astronomie der Universit\"{a}t Heidelberg, M\"{o}nchhofstra\ss e 12-14, 69120 Heidelberg, \\ Germany\\
$^5$Max-Planck Institut f\"{u}r Astronomie, K\"{o}nigstuhl 17, 69117 Heidelberg, Germany \\
}

\maketitle

%									Abstract & Key words								%
\begin{abstract}

The inner few hundred parsecs of the Milky Way harbours gas densities, pressures, velocity dispersions, an interstellar radiation field and a cosmic ray ionisation rate orders of magnitude higher than the disc; akin to the environment found in star-forming galaxies at high-redshift. Previous studies have shown that this region is forming stars at a rate per unit mass of dense gas which is at least an order of magnitude lower than in the disc, potentially violating theoretical predictions. We show that all observational star formation rate diagnostics -- both direct counting of young stellar objects and integrated light measurements -- are in agreement within a factor two, hence the low star formation rate is not the result of the systematic uncertainties that affect any one method. As these methods trace the star formation over different timescales, from $0.1-5$\,Myr, we conclude that the star formation rate has been constant to within a factor of a few within this time period. We investigate the progression of star formation within gravitationally bound clouds on $\sim$\,parsec scales and find $1-4$\,per cent of the cloud masses are converted into stars per free-fall time, consistent with a subset of the considered ``volumetric'' star formation models. However, discriminating between these models is obstructed by the current uncertainties on the input observables and, most importantly and urgently, by their dependence on ill-constrained free parameters. The lack of empirical constraints on these parameters therefore represents a key challenge in the further verification or falsification of current star formation theories.

\end{abstract}
\begin{keywords}
stars: formation$-$ISM: clouds$-$ISM: HII regions$-$Galaxy: centre.
\end{keywords}
\label{firstpage}

%%%%%%%%%%%%%%%%%%%%%%%%%%%%%%%%%%%%%%%%%%%%%%%%%%%%%%%

\section{Introduction}\label{intro}

%\vspace{5cm}

The inner few hundred parsecs of the Milky Way, known as the ``Central Molecular Zone" (CMZ), contains $\sim$\,80\,per cent of the Galaxy's dense molecular gas (2\,-\,6\,$\times10^{7}$\,\sol; \citealp{morris_1996}). The conditions within this region are extreme compared to those within the Galactic disc: the average density, temperature, pressure, velocity dispersion, interstellar radiation field and cosmic ray ionisation rate are factors of a few to several orders of magnitude larger. However, it has been noted for several decades that despite harbouring this vast reservoir of dense gas, the CMZ appears to be underproducing stars with respect to nearby star-forming regions in the disc (e.g. \citealp{guesten_1983, caswell_1983, taylor_1993, longmore_2013}). Understanding this dearth of star formation has wider implications, as the extreme properties of the CMZ are similar to those observed in the centres of nearby galaxies, starburst galaxies, and high redshift galaxies at the epoch of peak star formation density at z\,$\sim\,1-3$ \citep{kruijssen_2013}.

There could be several possible explanations for the apparent lack of star formation within the Galactic Centre, which can be split into two categories. Either the low star formation rate is a result of observational bias or uncertainty, or is the product of a physical mechanism.

The first observational explanation, could be that the gas is less dense than commonly assumed, and therefore should not form stars at such a high rate. The gas density is a difficult quantity to determine, as inferences of the three-dimensional structure have to be made in order to convert two-dimensional line-of-sight column density measurements. The measured column density of hydrogen in the CMZ appears to be at least an order of magnitude higher than clouds within the disc ($>\,10^{22}$\,cm$^{-2}$; \citealp{rathborne_2014}; Battersby et al. in prep), implying an average gas volume density above $\sim$\,$10^{4}$\,cm$^{-3}$. However, if the gas is more extended along the line of sight than in the plane of the sky, the density would be overestimated. Surveys using ATCA, APEX, and the SMA have shown that high critical density molecular gas tracers are widespread and spatially trace the peaks in column density (e.g. \citealp{jones_2012}). Additionally, these tracers have line of sight velocities which are consistent with being at the distance of the Galactic Centre (\citealp{ginsburg_2016, henshaw_2016}; Keto et al. in prep; Battersby et al. in prep). Recent radiative transfer modelling of the emission from dense molecular gas have shown that the gas has a density of the order $\sim\,10^{4}$\,cm$^{-3}$ \citep{Armijos-Abendano_2015, ginsburg_2016}. We conclude it is reasonable to assume that a significant fraction of the gas has a density $>\,10^{4}$\,cm$^{-3}$, and therefore remove this as a potential explanation for the apparent dearth of star formation. 

The second possible observational explanation for the apparent dearth in star formation is that methods to determine the star formation rate have systematic differences when applied to the CMZ compared to other environments. Star formation rates within local clouds are primarily determined by counting the embedded young-stellar population (YSO counting; refer to section\,\ref{YSO counting}). However, it is not possible to use this technique in external galaxies, as the individual sites of star formation cannot be resolved. Instead, the star formation rate is determined from integrated light measurements (e.g. infrared and free-free emission; refer to section\,\ref{free-free emission} and \ref{Infrared luminosities}). Our proximity to the centre of the Galaxy means that it is the only {\it extreme} environment in which comparison between YSO counting and integrated light measurement methods can be made. However, compared to the solar neighbourhood star-forming regions, the visual extinction is orders of magnitude higher (some positions have A$_\mathrm{V}>1000$\,mag), and contamination from non-associated (e.g. bulge) stars are more of an issue. This could result in systematic uncertainties in the YSO counting method. Furthermore, as we observe the Galactic Centre through the disc of the Milky Way, contamination of sources along the line-of-sight may also be an issue for the integrated light methods. A combination of these systematic uncertainties could lead to unreliable star formation estimates from any given method. 

The first physical explanation for the apparent lack of star formation within the Galactic Centre may be that star formation is episodic \citep{kruijssen_2014a}. \citet{krumholz_2015} and \citet{krumholz_2016} have modelled the dynamics of gas flows funnelled into the CMZ from large radii as acoustic instabilities within the bar's inner Lindblad resonance \citep{montenegro_1999}. In this model, when the gas reaches a radius of $\sim$\,100\,pc, and the rotation curve turns from flat to near-solid body, there is a decrease in shear which stops the inward flow and gas begins to accumulate. This accumulation of mass proceeds until the density is high enough for the gas to become gravitationally unstable, at which point there is an episode of intense star formation. The feedback from the recently formed high-mass stars then begins to drive turbulence and thereby increase the virial ratio of the gas, which quenches the star formation. Then as feedback from these stars' fades, gas can again accumulate and the cycle repeats. The estimated cycle timescale for a Milky Way-like galaxy is $\sim\,10-20$\,Myr. \citet{emsellem_2015} have conducted high-resolution, numerical simulations of the large scale gas motions within a galaxy similar to the Milky Way. These authors also find that gas is funnelled along the bar into the central $\sim$\,100\,pc, where transient star-forming complexes are observed, with timescales of a few Myr. \citet{torrey_2016} have tested the stability of feedback-regulated star formation for different environmental properties (e.g. ambient density, pressure) in the centres of galaxies. These authors find that a steady equilibrium state of star formation, where the energy input from feedback (which stops gravitational collapse) is balanced by the energy dissipation (which allows gravitational collapse), cannot be reached within the Galactic Centre, again requiring some degree of episodicity. \citet{suzuki_2015} also predict time-dependent flows, but these are instead driven via magnetic instabilities generated by differential rotation of the galaxy. Although these models and simulations differ in many aspects, the predicted trends in star formation activity are broadly similar in that they follow the scenario proposed by \citet{kruijssen_2014a}: gas steadily accumulates until a critical point is reached, when it becomes gravitationally unstable, collapses, and rapidly forms stars. Star formation continues until it is quenched by feedback, and the cycle restarts. 

\citet{leroy_2013} selected a sample of 30 nearby galaxies from the HERACLES survey to study the distribution of gas and stars on scales of $\sim$\,1\,kpc. They find a $\sim$\,1\,dex scatter on the gas depletion time (i.e. the time taken for all the gas to be converted to stars at the current star formation rate) towards the central $<0.5$\,kpc of the galaxies within their sample. This is a $\sim$\,0.3\,dex increase when compared to similar measurements in the disc of the same galaxies. This could be suggestive that episodic star formation is not limited to the centre of the Milky Way, but is also present within centres of other galaxies.

The second physical explanation for the apparent lack of observed star formation may be that the comparison to the predictions from star formation models may need revision (see section\,\ref{models} for discussion). These models have been benchmarked against regions in the solar neighbourhood (see \citealp{federrath_2012} and references within), so the predictions may not be directly applicable to extreme environments (e.g. as is found in the Galactic Centre). 

In this paper we investigate the three outstanding (observational and physical) explanations for the low star formation rate observed within the Galactic Centre: 1) inconsistent star formation rate measurements, 2) episodic star formation, 3) inappropriate comparison to the predictions of theoretical star formation models. To do this, we use infrared luminosities to determine the star formation rate over global and local (cloud) scales, and compare these to existing measurements and predictions from star formation models. In section\,\ref{method} we discuss how both the luminosities and column densities are determined. In section\,\ref{global} we determine the global star formation rate, compare this to previous measurements and predictions from star formation models, and discuss the implications. In section\,\ref{embedded} we determine the gas mass, embedded stellar mass and associated uncertainties of several sources within the Galactic Centre. In section\,\ref{sec:SFE} we determine the star formation efficiencies and star formation rates, and compare to the predictions of star formation models. In section\,\ref{conclusions} we discuss and summarise the implications of these results.

\section{Bolometric luminosity maps of the Galactic Centre}\label{method}

\begin{figure}
\centering
\includegraphics[trim = 0mm 6mm 0mm 0mm, clip,angle=0,width=1\columnwidth]{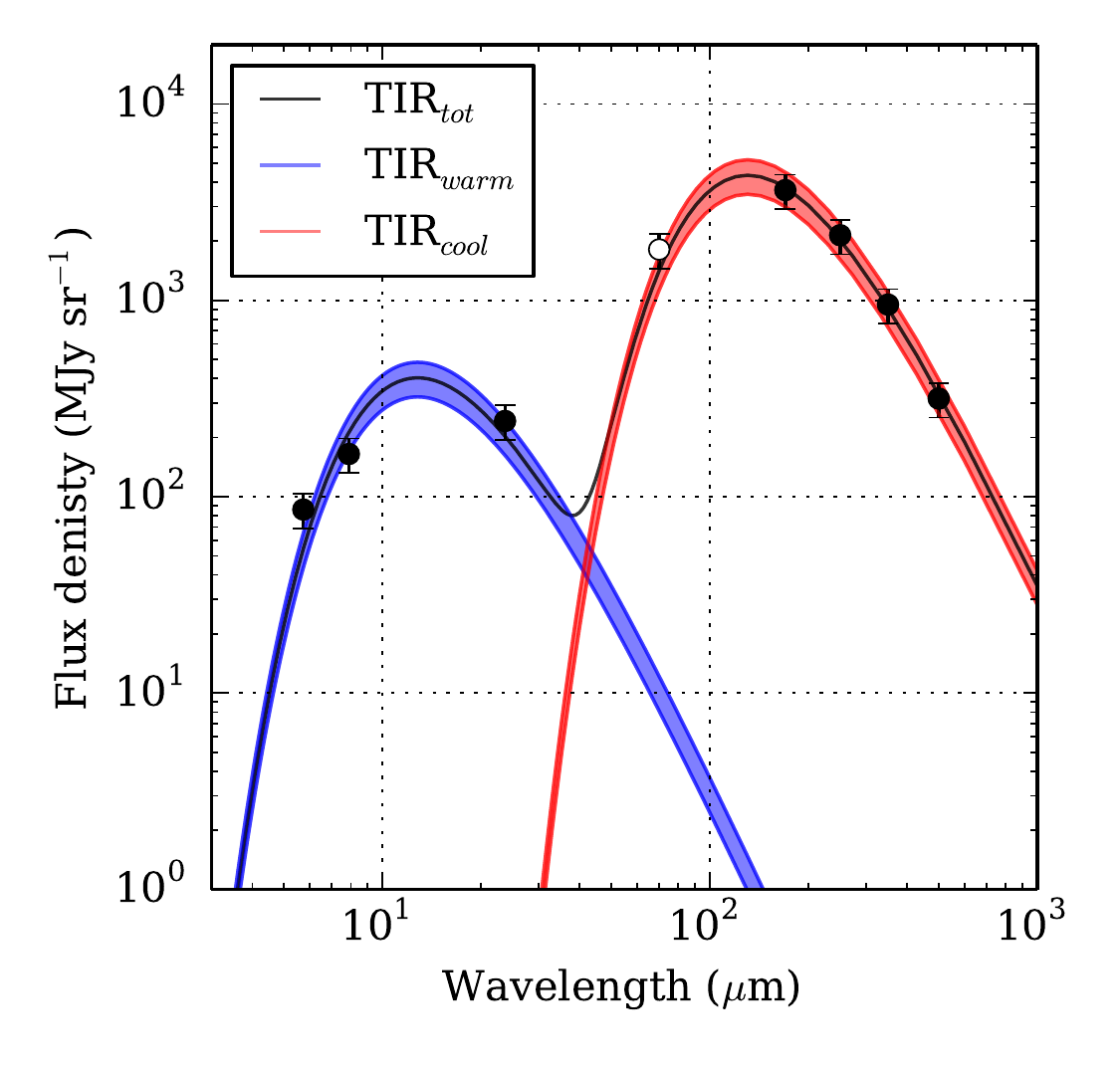}
\caption{The average background subtracted spectral energy distribution for the $|l|$$<$1$^{\circ}$, $|b|<$0.5$^{\circ}$ region (see Figure\,\ref{sfr_maps}). The points show the flux density for each wavelength (70\micron\ point is shown but is not used in the fitting), and the lines represent the warm component fit in dashed red (TIR$_{warm}$), the cool component in dashed blue (TIR$_{cool}$) and the total fit in solid black (TIR$_{tot}$\,=\,TIR$_{warm}$\,+\,TIR$_{cool}$). Error bars show the estimated $\sim$\,20\,per cent uncertainty on each point, and the shaded region represents the uncertainty on each fit.} 
\label{sed}
\end{figure}

\begin{figure*}
\centering
\includegraphics[trim = 0mm 0mm 0mm 0mm, clip,angle=0,width=1\textwidth]{\dir 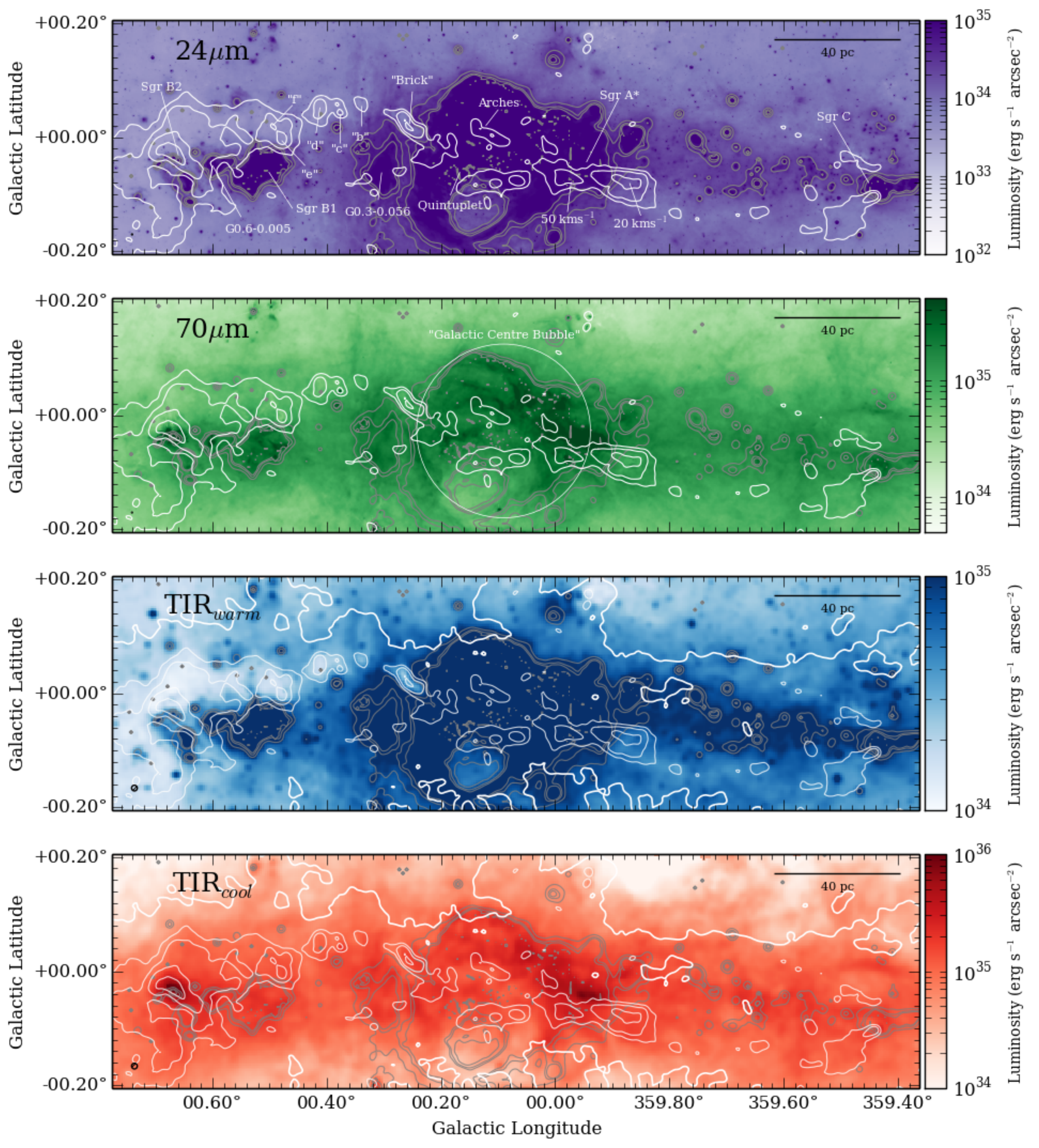}
\caption{Maps of the extinction-corrected 24\micron \ luminosity (upper panel, purple), 70\micron \ luminosity (upper middle panel, green), and warm (lower middle panel, blue) and cool (lower panel, red) components of the bolometric luminosity. Overlaid are the warm component column density contours in grey of \colhtwowarm\,=\,$\{1, 1.9, 2.6\}\times10^{17}$\,cm$^{-2}$, and cool component contours in white of \colhtwocool\,=\,$\{5, 10, 24, 75\}\times10^{22}$\,cm$^{-2}$. These contours levels are used to define the various sources (see Table\,\ref{masses_table}). The thick white contour shown in the lower two panels is of \colhtwocool\,=\,$1\times10^{22}$\,cm$^{-2}$. This contour highlights the widespread distribution of the cool column density component, which dominates the total column density \colhtwo\,=\,\colhtwowarm\,+\,\colhtwocool. The 24\micron\ luminosity map (upper panel) has several sources labeled, and the 70\micron\ map has the ``Galactic Centre Bubble'' shown (upper middle panel; e.g. \citealp{bally_2010}). Each panel has a scale bar located in the top right which represents 40\,pc at a distance of $\sim$\,8.5\,kpc \citep{reid_2014} and a circle in the bottom left which represents the beam size of the observations.} 
\label{sfr_maps}
\end{figure*}

\begin{table*}
\caption{Summary of survey data. Columns show the wavelength of the band, the beam full width at half maximum (FWHM), the image pixel size, the extinction with respect to the K-band, and the survey from which the observations were taken.}
\centering
\begin{tabular}{ c c c c c }
\hline
Band (\micron) & FWHM (\arcsec) & R$_{pix}$ (\arcsec) & A$_\lambda$/A$_k$ & Survey \\ \hline
5.8 & 2 & 1.2  & 0.44$^a$ & GLIMPSE$^c$ \\
8 & 2 & 1.2 &  0.43$^a$ & GLIMPSE$^c$ \\
24 & 6 & 2.4 &  0.61$^a$ & MIPSGAL$^d$ \\
70 & 5 &  3.2  & 0.06$^b$ & Hi-GAL$^e$ \\
160 & 11 & 4.5 & \dots & Hi-GAL$^e$ \\
250 & 18 & 6.0 & \dots & Hi-GAL$^e$ \\
350 & 25 & 8.0 & \dots & Hi-GAL$^e$ \\
500 & 36 & 11.5 & \dots & Hi-GAL$^e$ \\ \hline
\end{tabular}
\label{surveys}

{ \vspace{0.2cm}}
\begin{minipage}{\textwidth}
\vspace{1mm}
$^a$ Relations taken directly from \citet{chapman_2009}.\\
$^b$ Calculated using conversion from \citet{suutarinen_2013}.\\
$^c$ The {\it Spitzer} GLIMPSE is presented by \citet{churchwell_2009}.\\
$^d$ The {\it Spitzer} MIPSGAL survey is presented by \citet{carey_2009}.\\
$^e$ The {\it Herschel} Hi-GAL survey is presented by \citet{molinari_2010}.\\
\vspace{0mm}
\end{minipage}
\end{table*}

To create the infrared luminosity maps of the Galactic Centre needed to derive the star formation rates, we use {\it Spitzer} and {\it Herschel} telescope observations. The wavelengths and resolutions of these observations are presented in Table\,\ref{surveys}. To account for interstellar extinction, we adopt an average K-band extinction of A$_\mathrm{K}\sim2$\,mag from \citet{finger_1999}, \citet{dutra_2003} and \citet{schodel_2010}, who determined the extinction for several objects within the CMZ region. This is applied to the $5.8-70$\,\micron\ wavelength maps using the extinction relations presented by \citet[][see Table\,\ref{surveys}]{chapman_2009}. To apply the extinction to the 70\,-\,500\,\micron \ emission, we use the extinction relation A$_\mathrm{\lambda}$/A$_\mathrm{K}\,\sim\,0.0042\,(250/\lambda$\,[\micron])$^2$ \citep{suutarinen_2013}. From this, the 70\micron\ luminosity is corrected by A$_\mathrm{70\,\mu m}$\,=\,0.06\,mag, whilst for wavelengths larger than 70\micron\ the correction is negligible. To remove the background emission from the $70-500$\,\micron\ data, we follow the method outlined by \citet[][background removal of these data will be presented by Battersby et al. in prep]{battersby_2011}. 

Figure\,\ref{sed} displays the extinction corrected average spectral energy distribution for an example region ($|l|$$<$1$^{\circ}$, $|b|<$0.5$^{\circ}$). This shows that the Galactic Centre shows two distinct temperature components (see Figure\,\ref{sed}). The first peaks at wavelengths $\sim$\,$100-200$\,\micron, and is thought to originate from the cool dust (T\,$\sim$\,30\,K). The second peaks at $\sim$\,10\,\micron\ (T\,$\sim$\,300\,K), and becomes prominent towards known star-forming regions, and originates from warm dust. To measure the total column density of hydrogen (\colhtwo), the dust temperatures and the total bolometric luminosity, we fit a two component modified Planck function to the spectral energy distribution at each pixel (after smoothing all wavelengths to the largest resolution of $\sim$\,36\arcsec). The modified Planck function with respect to frequency, ${\rm S}_\nu$, takes the form, 
\begin{equation}
{\rm S}_\nu = \frac{2 h \nu^3}{c^2 \left(  e^{h\nu/kT} - 1 \right)} \left( 1 - e^{-\tau_\nu} \right), 
\end{equation}
where the opacity is given by, 
\begin{equation}
\tau_\nu = \mu_\mathrm{H_2} m_\mathrm{H} \kappa_\nu N_\mathrm{H}.
\end{equation}
We assume a mean molecular weight of $\mu_\mathrm{H_2}$=2.8\,a.m.u, a dust opacity of $\kappa_{\nu}\,=\,\kappa \left( \nu / \nu_{0} \right)^\beta$ at $\nu$\,=\,505\,GHz with a $\kappa_0$\,=\,4.0 and $\beta$\,=\,1.75 \citep{battersby_2011}, and a constant gas-to-dust ratio of 100. 

To separate the two temperature components, we consider wavelengths between 5.8\,-\,24\,\micron\ for the warm component, and 160\,-\,500\,\micron\ for the cool component (cool component is provided by Battersby et al. in prep). We do not consider {\it Spitzer} data with wavelengths smaller than 5.8\,\micron\ when fitting the spectral energy distribution, as it is not clear how much flux at these wavelengths is from recently formed stars or the older stellar population in the bulge of the Galaxy. Overlaid on Figure\,\ref{sed} are the average warm (TIR$_{warm}$) and cool component (TIR$_{cool}$) fits, and the total fit (TIR$_{tot}$\,=\,TIR$_{warm}$\,+\,TIR$_{cool}$). 

The shaded region for the average fits represents the instrumental uncertainty, which has been estimated as a conservative $\sim$\,20\,per cent on the flux density measurements. However, we expect the absolute uncertainty to be higher than this, due to variations in the dust properties across the region. For example, several authors have shown that there is a gradient of decreasing gas-to-dust ratio with decreasing galactocentric radius \citep{schlegal_1998, watson_2011}, a trend which has also been observed in several other star-forming galaxies \citep{sandstrom_2013}. Assuming that the gas-to-dust ratio is inversely proportional to the metallicity, the gas-to-dust ratio within the central kpc of the Galaxy would be $\sim$\,50 (e.g. \citealp{sodroski_1995}), which would cause the column densities, and gas mass measurements later in this work (see section\,\ref{embedded}), to be a factor of two lower. Given this, we estimate the absolute column density measurements should be reliable to within a factor of two. 

The flux-densities, ${\rm S}_\nu$, are converted into fluxes, S (in units of MJy\,sr$^{-1}$), by integrating the two component modified black body at each position. To convert these into luminosities (units of erg\,s$^{-1}$ or \Lsol) requires an accurate measurement of the source distance. This analysis, therefore, has been restricted to the $|l|$$<$1$^{\circ}$, $|b|<$0.5$^{\circ}$ region, as parallax measurements \citep{reid_2009, reid_2014} and modelling \citep{molinari_2011, kruijssen_2015} have shown that the majority of the dense molecular gas and star-forming regions are close to the Galactic Centre. Additionally, the extreme environment within this region has many identifiable features (e.g. large velocity dispersion), which have been used to show that there is little contamination from non-associated material along the line-of-sight (e.g. \citealp{henshaw_2016}). We are, therefore, confident with the distance measurement to the $|l|$$<$1$^{\circ}$, $|b|<$0.5$^{\circ}$ region of $\sim$\,8.5\,kpc \citep{reid_2014}. The integrated flux, S (MJy\,sr$^{-1}$), is converted to luminosity, L (erg\,s$^{-1}$), with the units shown in parenthesis using, 
\begin{equation}
{\rm L} ({\rm erg\,s}^{-1}) = 2.8\,\times\,10^{10}\,{\rm S}_\nu({\rm MJy\,sr}^{-1})\,{\rm R}_{pix}^{2}(^{\prime\prime})\,{\rm D}^{2}({\rm pc}),
\end{equation}
where R$_{pix}$ is the pixel size and D is the distance to the region. Figure\,\ref{sfr_maps} presents the 24\micron, 70\micron, and the warm and cool component bolometric luminosity maps (TIR$_{warm}$ and TIR$_{cool}$, respectively) for the $|l|<$1$^{\circ}$, $|b|<$0.5$^{\circ}$ region. Over-plotted are grey and white contours of the warm and cool gas column densities, respectively. Labels shows the positions of the main objects of interest.  The total luminosities within $|l|$$<$1$^{\circ}$, $|b|<$0.5$^{\circ}$ are L(24\micron)\,=\,9.4\,$\pm$\,1.9\,$\times\,10^{7}$\,\Lsol, L(70\micron)\,=\,3.4\,$\pm$\,0.7\,$\times\,10^{8}$\,\Lsol, L(TIR$_{tot}$)\,=\,5.7\,$\pm$\,1.7\,$\times\,10^{8}$\,\Lsol. 

Throughout this work, we make the standard assumption that all the emission from the embedded stellar population is reprocessed by the surrounding dust to infrared wavelengths, which is emitted at much shorter wavelengths at the interface where the dust becomes optically thin. In this scenario, the total infrared luminosity directly corresponds to the bolometric luminosity produced by the embedded population. We estimate the measurement uncertainty on the total bolometric luminosity as the maximum variation after changing the flux densities at each wavelength by $\pm$\,20\,per cent (an upper limit estimate of the flux uncertainties) and re-fitting the spectral energy distribution. These uncertainties are, however, small when compared to the systematic uncertainties, for example: i) leakage of high-energy photons (in which case the infrared luminosity is not equal to the total bolometric luminosity), ii) heating of dust via other sources, and iii) emission produced from the older embedded population within the CMZ (e.g. \citealp{calzetti_2010}). It is difficult to estimate the amount of energy leakage, as this requires an accurate description of the three-dimensional density structure of the individual star-forming regions. Similarly, for the dust heating from other sources, such as from the central super massive black hole (Sgr A$^*$), it is difficult to estimate given our limited knowledge of the radiation field within this region. However, an estimate of the contribution from the older embedded population can be made using the Besan\c{c}on model \citep{robin_2003}. To estimate the expected bolometric luminosity from the field stellar population we use the online\footnote{\url{http://model.obs-besancon.fr}} catalogue simulation, which includes all the stellar luminosity classes and ages. Within the region $|l|$$<$1$^{\circ}$, $|b|<$0.5$^{\circ}$, taking all stars with a distance between 8.4\,-\,8.6\,kpc (i.e. forming a $\sim$\,200$\times$100$\times$200\,pc box containing the CMZ), we find that the total bolometric luminosity from the population older than 0.15\,Gyr is $\sim\,2.5\times\,10^{8}$\,L$_{\odot}.$\footnote{We note that \citet{robin_2003} do not fit the model to the observed stellar density within $|l|$$<$1$^{\circ}$. Rather the stellar densities are predicted by extrapolating a power-law, from the DENIS survey within $-8^{\circ}<l<12^{\circ}$, $|b|<$4$^{\circ}$ (see \citealp{epchtein_1997}).}

The \citet{robin_2003} bulge population luminosity of $\sim\,2.5\times\,10^{8}$\,L$_{\odot}$ is similar to that determined by \citet[][see their Table 6, 1.9\,$\times\,10^{8}$\,\Lsol]{launhardt_2002}, from $2.2-240$\micron\ IRAS and COBE data. The \citet{robin_2003} and \citet{launhardt_2002} estimates are around half of the total measured luminosity we find within $|l|$$<$1$^{\circ}$, $|b|<$0.5$^{\circ}$ (5.7\,$\pm$\,1.7\,$\times\,10^{8}$\,\Lsol). However, the spectral energy distribution of old stars will peak at wavelengths $<5$\,\micron. To estimate the direct contribution of the old stellar population to the infrared luminosity in the wavelength range between $5.8-500$\,\micron\ we integrate the black-body spectral energy distribution with the average effective temperature of all stars from the \citet{robin_2003} stellar population model ($\sim$\,3000\,K). We find that the fraction of the luminosity produced by these stars emitted between $5.8-500$\,\micron\ is $\sim$\,3\,per cent. This suggests that approximately $\sim$\,$1-2$\,per cent of the total infrared luminosity within $|l|$$<$1$^{\circ}$, $|b|<$0.5$^{\circ}$ is directly produced by the old bulge star population. We, therefore, do not remove this contribution from the luminosity and conclude the bolometric luminosity between $5.8-500$\,\micron\ is dominated by the emission from young stars.

%%%%%%%%%%%%%%%%%%%%%%%%%%%%%%%%%%%%%%%%%%%%%%%%%%%%%%%%
%										Results										%
%%%%%%%%%%%%%%%%%%%%%%%%%%%%%%%%%%%%%%%%%%%%%%%%%%%%%%%%

\section{Global star formation}\label{global}

\subsection{Determining the global star formation rate}\label{totalsfr}

In order to estimate the total star formation rate across the CMZ, we apply several infrared luminosity-to-star formation rate (luminosity-SFR) relations to the bolometric and monochromatic infrared luminosities (see \citealp{kennicutt_2012} and references therein). These relations are based on the assumption that high-mass ``young stellar objects'' (YSOs) classified as having ages in the range $\sim$\,$0.1 - 5$\,Myr (see section\,\ref{litsfr}), are still heavily embedded within their parent molecular clouds when they first reach the zero age main sequence (ZAMS). Therefore, the majority of their prodigious short wavelength (e.g. ultra-violet) emission is absorbed by dust within their surrounding medium, and re-emitted at longer infrared wavelengths ($\sim$\,$1-1000$\micron). Hence, the infrared luminosity can be used, similar to a calorimeter, to estimate the underlying embedded population. Given that high-mass stars have a characteristic age of a few Myr, the star formation rate can then be estimated. One advantage of this method is that it does not require the individual sites of star formation to be resolved ($<$\,0.05\,pc). By using the integrated luminosity of an entire stellar population, over scales of $\goa\,$100\,pc, the luminosity-SFR relations can be used to determine the star formation rates within extragalactic sources for which it is impossible to resolve individual forming stars. A sample of the most widely used monochromatic and bolometric luminosity-SFR relations are summarised in Table\,\ref{SFR_conv}, with the luminosity limits over which they are considered to be reliable. Table\,\ref{SFR_conv} also shows the global star formation rates within the region $|l|$$<$1$^{\circ}$, $|b|<$0.5$^{\circ}$ derived using these relations. We find that the average star formation rates derived from the 24\micron, 70\micron, and TIR luminosities are 0.09$\pm$\,0.02, 0.10$\pm$\,0.02, and 0.09$\pm$\,0.03\,\solyr, respectively. The uncertainties shown here are from the measurement uncertainties on the luminosity (see section\,\ref{method}). We note, however, that the systematic uncertainty on the luminosities are significantly larger, and the luminosity-SFR relations have an uncertainty of around a factor two. Taking these uncertainties into account, we estimate that the measured star formation rates are reliable to within a factor of two. Hence the average {\it global} star formation rate within the $|l|$$<$1$^{\circ}$, $|b|<$0.5$^{\circ}$ region derived from the luminosity-SFR relations has a mean of $0.09\pm0.02\,$\solyr.

\begin{table*}
\caption{Summary of various SFR conversion factors. Columns show the wavelengths to which the conversions apply, the conversion reference, the conversion factors (where L($\lambda$) is the luminosity at wavelength $\lambda$), the reliability limits, and the calculated star formation rate within within $|l|$$<$1$^{\circ}$, $|b|<$0.5$^{\circ}$.}

\centering
\begin{tabular}{ c c c c c}
\hline
	Wavelengths for & Conversion & Luminosity range$^{a}$ & Reference for & SFR within $|l|$$<$1$^{\circ}$, $|b|<$0.5$^{\circ}$ \\ 
	conversion (\micron) & (erg s$^{-1}$ \sol$^{-1}$ yr) & (erg\,s$^{-1}$) & conversion & (\sol yr$^{-1}$) \\ \hline
	
	24		& 2.75\,$\times\,10^{-43}$\,L(24\micron) & (0.01\,-\,1)\,$\times\,10^{44}$ & \citet{wu_2005} & 0.10\\ 
	 	 	& 2.46\,$\times\,10^{-43}$\,L(24\micron) & (0.004\,-\,2)\,$\times\,10^{44}$ & \citet{zhu_2008} & 0.09\\ 
	  		& 2.04\,$\times\,10^{-43}$\,L(24\micron) & (0.4\,-\,5)\,$\times\,10^{43}$ & \citet{rieke_2009}  & 0.07$^{b}$\\ 
			& 9.01\,$\times\,10^{-34}$\,L(24\micron)$^{0.768}$ & (0.001\,-\,3)\,$\times\,10^{41}$ & \citet{perez-gonzalez_2006} & 0.07\\
			& 5.83\,$\times\,10^{-38}$\,L(24\micron)$^{0.871}$ & (0.0001\,-\,3)\,$\times\,10^{44}$ & \citet{alonso-herrero_2006} & 0.09\\
			& 1.31\,$\times\,10^{-38}$\,L(24\micron)$^{0.885}$ & (0.0001\,-\,3)\,$\times\,10^{44}$ & \citet{calzetti_2007} & 0.08\\
			& 5.66\,$\times\,10^{-36}$\,L(24\micron)$^{0.826}$ & (0.000001\,-\,3)\,$\times\,10^{44}$ & \citet{relano_2007} & 0.12 \\ \hline
			
	70 		 & 5.88\,$\times\,10^{-44}$\,L(70\micron) & $>$\,1.4\,$\times\,10^{42}$ & \citet{calzetti_2010} & 0.07\\ 
			 & 9.37\,$\times\,10^{-44}$\,L(70\micron) & (0.005\,-\,5)\,$\times\,10^{43}$ & \citet{li_2010}  & 0.12\\ 
			 & 9.70\,$\times\,10^{-44}$\,L(70\micron) & \dots & \citet{lawton_2010} & 0.12\\ \hline
			 
	TIR 		& 4.5\,$\times\,10^{-44}$\,L(TIR) & \dots & \citet{kennicutt_1998} & 0.10$^{c}$\\
			& 3.88\,$\times\,10^{-44}$\,L(TIR) & (0.02\,-\,2)\,$\times\,10^{43}$ & \citet{murphy_2011} & 0.09\\ \hline

\end{tabular}
\label{SFR_conv}
{ \vspace{0.4cm}}
\begin{minipage}{\textwidth}
\vspace{2mm}
$^a$ Note, most of the authors do not specify a luminosity range of the validity of the SFR conversion. Therefore, following \citet{calzetti_2010}, we define the luminosity range as the limits of the sample in each work. \\
$^b$ \citet{yusef_2009} estimate a star formation rate of 0.07\solyr, using this conversion with a 24\,\micron\ luminosity over an area of $|l|$$<$1.3$^{\circ}$, $|b|<$0.17$^{\circ}$.\\
$^c$ \citet{crocker_2011} estimate a star formation rate of 0.08\solyr, using this conversion with a bolometric luminosity found from 2.2\,-\,240\micron \ IRAS data.
\end{minipage}
\end{table*}

\subsection{Comparison to star formation rates within the literature}\label{litsfr}

In this section, we discuss the methods which have been previously used to determine the total star formation rate within the Galactic Centre, and compare to the values determined in section\,\ref{totalsfr}.

\subsubsection{YSO counting}\label{YSO counting}

The first method uses infrared emission to measure the masses of individual high-mass YSOs (ages $\sim$\,$0.1-1$\,Myr; M\,$\ga$\,10\,\sol). As previously mentioned, high-mass stars reach the ZAMS whilst they are still heavily embedded within their parent molecular cloud (and are most likely still accreting material). They inject a significant amount of energy into their surrounding environment, and can therefore be identified from their strong, compact near/mid infrared emission. Once identified, their masses can be estimated from their bolometric luminosity. The total embedded stellar population mass of a region can then be inferred by extrapolating the stellar initial mass function down to lower masses assuming an appropriate initial mass function (IMF). Using this method, \citet{yusef_2009} identified potential YSOs as sources which show excess 24\micron\ emission with respect to 8\micron\ emission within the region $|l|<1.3^{\circ}$, $|b|<0.17^{\circ}$. By modelling the spectral energy distributions, they distinguish which are the young sources, and measure their luminosities to estimate the masses. They find the total embedded stellar population after IMF extrapolation is $\sim$\,1.4\,$\times\,10^{4}$\,\sol. These authors assume a YSO lifetime of 0.1\,Myr, which they use to estimate a global star formation rate of $\sim\,0.14$\,\solyr\ ($|l|<1.3^{\circ}$, $|b|<0.17^{\circ}$). 

There are several inherent difficulties with the YSO counting method. For example, it is not trivial to determine the YSO ages from either the infrared excess, or the spectral energy distributions. Furthermore, dusty, bright, asymptotic giant branch stars are also known to emit at infrared wavelengths with similar colours to YSOs \citep{habing_1996}. Therefore, YSO identification can be plagued with source contamination. In light of this, \citet{koepferl_2015} re-examined the \citet{yusef_2009} YSO sample, comparing in-depth radiative transfer modelling of both embedded YSOs and embedded older main sequence stars (ages of $>$\,1\,Myr). These authors conclude the \citet{yusef_2009} sample suffers from significant contamination from embedded main sequence stars (which produce very similar emission profiles within the wavelengths used by \citealp{yusef_2009}). They propose the star formation rate is factors of several lower: $\sim\,0.06$\,\solyr. \citet{immer_2012} used infrared spectral features to attempt to disentangle the young and more evolved stars. These authors initially identified candidate YSOs as compact sources with excess 7\micron/15\micron\, emission, which are verified by examining the infrared spectral features of a test sample. Within $|l|$$<$1.5$^{\circ}$, $|b|<$0.5$^{\circ}$, \citet{immer_2012} estimate a total embedded stellar population mass of $\sim$\,7.7\,$\times\,10^{4}$\,\sol. Given the slightly different selection criteria used by \citet{immer_2012}, these authors assume a longer YSO lifetime of $\sim$\,1\,Myr compared to \citet{yusef_2009} and \citet{koepferl_2015}, yet calculate a comparable average SFR of $\sim\,0.08$\,\solyr.

\subsubsection{Free-free emission}\label{free-free emission}
The second method to determine star formation rates involves using cm-(mm-)-continuum emission to measure the mass of the YSO population. Along with heating the surrounding environment, YSOs with ages of $\sim$\,3\,Myr and masses $>$\,$8-10$\,\sol\ emit a significant amount of high-energy ionising radiation (i.e. photons with $h \nu$$>$13.6\,eV), which produces \HII\ regions. The free-free emission from the ionised gas (i.e. bremsstrahlung radiation) can be observed at cm-wavelengths, when the medium is optically thin. Cm-continuum emission observations, therefore, provide a reliable way to determine the rate at which ionising photons are produced from massive stars within a region, which in turn can be used to estimate their mass. The total embedded stellar population can then be extrapolated using an IMF, from which the star formation rate can be estimated (e.g. \citealp{murray_2010}). \citet{lee_2012} used {\it Wilkinson Microwave Anisotropy Probe (WMAP)} continuum observations at wavelengths of $\sim$\,$3-150$\,mm, to identify and measure sources of free-free emission within the Milky Way. \citet{longmore_2013} used this catalogue and the free-free-to-SFR conversion presented by \citet{murray_2010}, to estimate a global star formation rate of $\sim\,0.06$\,\solyr\ within the region $|l|$$<$1$^{\circ}$, $|b|<$0.5$^{\circ}$ (the same region adopted by this work). 

\subsubsection{Infrared luminosities}\label{Infrared luminosities}
The third method to estimate star formation rates involves using the bulk infrared emission and the luminosity-SFR relations. This method follows a similar basis to the YSO counting method, where the near/mid infrared dust emission is modelled to determine the embedded population. The key difference is that the luminosity-SFR relations use the integrated emission from entire stellar populations, hence sample star formation rates over larger times, which usually translates to larger spatial scales, than to YSOs counting (see Table\,\ref{areas}).

\citet{launhardt_2002} fit the spectral energy distribution of 2.2\,-\,240\micron\ IRAS and COBE data, and estimate the total infrared bolometric luminosity within $|l|<0.8^{\circ}$, $|b|<0.3^{\circ}$ is L(TIR$_{tot}$)\,=\,4.2$\times\,10^{8}$\,\Lsol. \citet{crocker_2011} use this, with the relation of \citet[][ see Table\,\ref{SFR_conv}]{kennicutt_1998} to calculate a SFR of $\sim$\,0.08\solyr. \citet{yusef_2009} have used the 24\micron\ monochromatic luminosity from {\it Spitzer}, of L(24\micron)\,=\,9\,$\times\,10^{7}$\,\Lsol\ within $|l|$$<$1.3$^{\circ}$, $|b|<$0.17$^{\circ}$, with the luminosity-SFR relation presented by \citet[][ also see Table\,\ref{SFR_conv}]{rieke_2009}, to estimate a star formation rate of $\sim$\,0.07\,\solyr.

\subsubsection{Comparison of star formation rates}
In Table\,\ref{areas} we tabulate the measurements of the star formation rate within the Galactic Centre. Apart from the YSO counting measurement of \citet{yusef_2009}\footnote{This likely suffers from contamination, see section\,\ref{YSO counting}.}, we find agreement to within $\sim$\,35\,per cent between the various measurements. To attempt a fairer comparison between these star formation rates and those found in this work, we re-determine the star formation rates within the areas adopted by the works listed in Table\,\ref{areas}, when using the total bolometric luminosity and using a single luminosity-SFR relation \citep{kennicutt_1998}. These measurements are shown in Table\,\ref{areas}. As the total bolometric luminosity measurements are self-consistent, they exclusively reflect the effect of changing the considered area. Given this, we can conclude that the there is no systematic uncertainty in any one measurement method which is causing an underestimation of the star formation rate.

%This can also be shown by choosing a fiducial area (e.g. $|l|<1.3^{\circ}$, $|b|<0.17^{\circ}$) and rescaling the observed star formation rates listed in Table\,\ref{areas} to the same area using total bolometric luminosity derived star-formation rates, which gives values of 0.14, 0.06, 0.07, 0.08, 0.09, 0.06\,\solyr\ (in the order presented in Table\,\ref{areas}). Again, omitting the \citet{yusef_2009} value, we find when choosing the same area, each of the methods is consistent to within $\pm$\,0.02\solyr. 

\begin{table*}
\caption{Summary of star formation rate measurements within the literature. Shown is the measurement method, the characteristic age range traced by the method, the CMZ area over which the star formation rate has been measured, the measured star formation rate, the star formation rate determined using the infrared bolometric luminosity within the different areas of the CMZ and the luminosity-star formation-rate relation from \citep{kennicutt_1998}.}
\centering
\begin{tabular}{c c c c c c}
\hline
Method used to & Characteristic & Area of CMZ over & SFR determined from & SFR determined using the infrared \\
determine the SFR & age probed by & which the SFR is & the corresponding method & bolometric luminosity with the \\ 
 & method (Myr) &  calculated & and area (\solyr) & \citet{kennicutt_1998} relation from the \\ 
 & & & & corresponding area (\solyr) \\ \hline

YSO counting & $\sim$\,0.1 & $|l|<1.3^{\circ}$, $|b|<0.17^{\circ}$ & 0.14$^{b}$  & 0.07 \\
YSO counting & $\sim$\,0.1 & $|l|<1.3^{\circ}$, $|b|<0.17^{\circ}$ & 0.06$^{c}$  & 0.07\\
Infrared luminosities & 0-5-100$^{a}$ & $|l|$$<$1.3$^{\circ}$, $|b|<$0.17$^{\circ}$ & 0.07$^{b}$  & 0.07 \\
Infrared luminosities & 0-5-100$^{a}$ & $|l|$$<$0.8$^{\circ}$, $|b|<$0.3$^{\circ}$ & 0.08$^{d}$  & 0.07\\
YSO counting & $\sim$\,1 & $|l|$$<$1.5$^{\circ}$, $|b|<$0.5$^{\circ}$ & 0.08$^{e}$  & 0.12 & \\
free-free emission & $\sim$\,0-3-10$^{a}$ & $|l|$$<$1$^{\circ}$, $|b|<$0.5$^{\circ}$ & 0.06$^{f}$  & 0.10 & \\
\hline
\end{tabular}
\label{areas}
{ \vspace{0.4cm}}
\begin{minipage}{\textwidth}
\vspace{2mm}
$^{a}$ The second number indicates the mean age of the stellar population contributing to the emission, the third number shows the age below which 90\,per cent of emission is contributed \citep{kennicutt_2012}. \\
$^{b}$ \citet{yusef_2009}. \\
$^{c}$ \cite{koepferl_2015}. \\ 
$^{d}$ \citet{crocker_2011}. \\
$^{e}$ \citet{immer_2012}. \\
$^{g}$ \citet{longmore_2013}. \vspace{-3mm}
\end{minipage}
\end{table*}

\subsection{Comparison to star formation rates predicted from theoretical models}\label{models}

Given that we now have a set of consistent measurements for the global star formation rate within the Galactic Centre, comparison can be made to different star formation theories within the literature. Specifically, we focus on a ``column density'' threshold relation, a relation between the gas mass and the star formation rate, and three theoretical models based on ``volumetric'' gas density scalings.

\vspace{3mm}

We firstly consider the ``column density'' threshold relation presented by \citet{lada_2010, lada_2012}. \citet{lada_2010} found that, for local clouds, a correlation exists between the gas mass at high extinctions (A$_K$\,$>$\,0.8\,mag) and the number of embedded YSOs identified in the infrared. These authors estimate that this gas has a hydrogen column density of the order 6.7\,$\times10^{21}$\,cm$^{-2}$ (or $\sim\,10^4$\,cm$^{-3}$; assuming no line of sight contamination, and a typical spherical core of radius $\sim$\,0.1\,pc). They measure the star formation rates of clouds from the number of embedded YSOs, assuming an initial mass function median mass ($\sim$\,0.5\,\sol), and a median age spread of the clouds ($t_{\rm age}\sim$\,2\,Myr), such that SFR(\solyr)~=~N$_{\rm YSO}$\,M$_{{\rm IMF},median}$(\sol)\,/\,$t_{\rm age}$(yr). \citet{lada_2010} (and \citealp{lada_2012} who incorporate extragalactic sources) show that the amount of dense gas and the level of star formation are correlated. The measured depletion time of the dense gas is 20\,Myr, which implies,
\begin{equation}
\mathrm{SFR} = 
  \begin{cases} 
   0 & \text{if } N_\mathrm{H_2} < 6.7\times10^{21}\,\mathrm{cm}^{-2} \\
   4.6\times10^{-8}\,\mathrm{M_{gas}}  & \text{if } N_\mathrm{H_2} \geq 6.7\times10^{21}\,\mathrm{cm}^{-2}
  \end{cases}
\label{lada}
\end{equation}
where the star formation rate is in units of \solyr, and mass is in units of \sol.

%\vspace{2cm}

%\footnote{{\bf Observations have shown that at higher densities, the log-normal probability distribution turns to a power-law distribution, which suggest gravity is dominant (see \citealp{federrath_2013}, and references therein). This power-law distribution is, however, not described by any of the volumetric models reviewed by \citet{padoan_2014} and \citet{federrath_2012}.}}}

We now consider the ``volumetric'' models for star formation (see \citealp{padoan_2014} for a comprehensive review).  These are based on the assumption that the density distribution of the gas in star-forming regions follows a log-normal probability distribution when turbulence is dominant. These models use the dimensionless star formation efficiency per free-fall time, $\epsilon_\mathrm{ff}$, to describe the level of star formation within a region. This can be expressed as the integral of the probability distribution function, $p(x = \rho/\bar{\rho})$, above the critical overdensity of collapse, $x_{crit}=\rho_{crit}/\bar{\rho}$, where $\bar{\rho}$ is the mean density and $\rho_{crit}$ is the density when gravitational collapse begins to dominate. This integral is given as,
\begin{equation}
\epsilon_{\rm ff} = \frac{\epsilon_{core}}{\phi_{t}}\int_{x_{crit}}^{\infty} \frac{t_{\rm ff}(\bar{\rho})}{t_{\rm ff}(\rho)}\,p(x)\,dx,
\label{pdf}
\end{equation}
where $\epsilon_{core}$ is the fraction of core mass which forms the protostar, and $\phi_{t}$ is the gas replenishment factor, such that $\phi_{t}\,t_\mathrm{ff}$ is the replenishment time \citep{krumholz_2005}. There are two physical interpretations of this integral, which differ in the treatment of the density dependence of the free-fall time ($t_{\rm ff}(\bar{\rho})/t_{\rm ff}(\rho)$). Assuming a constant free-fall time based on the mean properties of the cloud simplifies the integral, and gives a ``single-free-fall'' solution where all the gas collapses over the same timescale ($t_{\rm ff}(\bar{\rho})/t_{\rm ff}(\rho)$ = constant). On the other hand, assuming that the free-fall time varies as a function of density, as suggested initially by \citet{hennebelle_2011}, requires the integral to be solved over the log-normal probability distribution. In this ``multi-free-fall'' time solution, smaller, denser structures can decouple from their lower density environment, and collapse on shorter timescales than the global free-fall time ($t_{\rm ff}(\rho>\bar{\rho})\,<\,t_{\rm ff}(\bar{\rho})$). This may be more representative of the hierarchical structure observed in star-forming regions.

The solutions to the integral shown in equation\,\ref{pdf}, require the calculation of the dispersion of the density probability distribution function with respect to the mean density, $\sigma_{x}$ (see \citealp{federrath_2012}). This can be estimated as,
\begin{equation}
\sigma_{x}\,\approx\,b\,\mathcal{M}\,(1\,+\,\beta^{-1})^{-1/2},
\label{sigma}
\end{equation}
where $b$ is the turbulence driving parameter. The turbulence driving parameter was introduced by \citet{federrath_2008} to distinguish solenoidal (divergence-free, $b\,\sim\,0.33$) driving of the turbulence from compressive (curl-free, $b\,\sim\,1$) driving. The turbulence driving varies from as low as $b\,=\,0.22\,\pm\,0.12$ in the ``Brick'' \citep{federrath_2016} to typical values of $b\,\approx\,0.5$ in several clouds in the Galactic disc (Taurus, IC5146, GRSMC43.30-0.33; see \citealp{federrath_2016}). In equation\,\ref{sigma}, $\mathcal{M}$ is the three-dimensional turbulent sonic Mach number,\footnote{The three-dimensional Mach number, denoted by $\mathcal{M}$, is used throughout this work.} and $\beta=2\mathcal{M}^2_\mathrm{A}/\mathcal{M}^2$ is used to quantify the importance of the magnetic field within the plasma, where $\mathcal{M}_\mathrm{A}$ is the Alfv{\'e}nic Mach number. Strong magnetic fields are, therefore, represented by low values of $\beta$, and a hydrodynamical (rather than magnetohydrodynamic) expression of equation\,\ref{sigma} can be obtained by setting $\beta\rightarrow\infty$. Shown in Table\,1 of \citet{federrath_2012} are the several definitions for the critical density for collapse, $x_{crit}$, and resultant solution to the integral of the density probability distribution function (equation\,\ref{pdf}). %Each derivation relies on a different fiducial factor, $\phi\approx1.12$ \citep[KM05]{krumholz_2005}, $\theta\approx0.35$ \citep[PN11]{padoan_2011}, and $y_\mathrm{cut}\approx0.1$ \citep[HC13]{hennebelle_2013}, and each has a solution assuming a single-free-fall time, derived from the global properties of the cloud, and multiple free-fall times (``multi-free-fall'), derived for each density in the probability distribution function to account for the range of local densities and free-fall times.}

\citet{longmore_2013} use two models to determine the predicted {\it global} star formation rate within the CMZ ($|l|$$<$1$^{\circ}$, $|b|<$0.5$^{\circ}$): the column density threshold relation of \citet{lada_2010}, as described above, and the model of \citet{krumholz_2012}, which is an evolution of the \citet{krumholz_2005} model. For the column density limit model, they estimate that $\sim$\,95\,per cent of the gas within the CMZ lies above the threshold for collapse, which gives a predicted star formation rate of 0.78\,\solyr. For the volumetric model, they then calculate the volume density by assuming that the gas at $|l|$$<$1$^{\circ}$ deg is distributed in a ring-like stream with a radius of $\sim$\,100\,pc \citep{molinari_2011}. This structure has a mass of 1.8$\,\times\,10^7$\sol, which gives a predicted star formation rate of 0.4\,\solyr. These predictions are significantly higher than the average observed star formation rate of $\sim$\,$0.08$\,\solyr. Table\,\ref{areas} shows the conclusion of \citet{longmore_2013} holds no matter which method is used to determine the star formation rate.

%\footnote{{\bf It has been suggested that multi-free-fall volumetric models provide predictions in better agreement with the observed global star formation rate within the CMZ \citep{federrath_2013a, salim_2015}.}}

\subsection{Implications for the global star formation rate}\label{impsfr}

In summary, we find that the star formation rates for the CMZ measured from the infrared luminosity, YSO counting, and free-free emission have a mean value across all measurements of 0.09\,$\pm$\,0.02\,\solyr, and given their uncertainties are in agreement to within a factor two. Furthermore, in agreement with the conclusion of \citet{longmore_2013}, we find that this is factors of a few to more than an order of magnitude smaller than is predicted from star formation models. In section\,\ref{intro} we speculated three possible causes for this apparent dearth in star formation within the Galactic Centre: 1) inaccurate star formation rate measurements, 2) episodic star formation, or 3) inappropriate comparison to the predictions from star formation relations/models. The results found in this section have shown:

\begin{itemize} 
\item[i)] The star formation rates determined from the infrared luminosity-SFR relations are within a factor two of previous measurements. This allows us to rule out that systematic uncertainties in the measurements are causing the apparent low star formation rate, unless this uncertainty affects all the methods in the same way, which seems unlikely. 

\item[ii)] The luminosity-SFR relations, which use the integrated light from the whole stellar population, and YSO counting methods, which require the sites of star formation to be resolved, are consistent in the Galactic Centre. As the Galactic Centre is the most extreme environment for which is it possible to resolve individual forming stars and make this measurement, the results here provide confidence that the luminosity-SFR relations reliably trace the star formation rate over kpc scales within similar environments present in starburst galaxies, and high-redshift galaxies.

\item[iii)] The various methods to determine the star formation rates are in agreement, despite being sensitive to star formation over different time scales over the past few Myr. Therefore, the global star formation rate has not changed over this time by more than a factor of two to three of its current rate. This is consistent with recent theoretical models predicting that the star formation rate in the CMZ is episodic on a timescale of $\sim$\,$10 - 20$\,Myr, much longer than the mean timescales covered by the adopted star formation rate tracers of $\sim$\,$0.1 - 5$\,Myr.

\end{itemize} 

Returning to the discussion in the introduction, we have ruled out the first of the possibilities for the apparent low star formation rate within the Galactic Centre, that it results from inaccurate star formation rate measurements. Furthermore, in agreement with recent theoretical work, we find that the Galactic Centre could be in a low point in a star formation cycle. \citet{kruijssen_2014a} have suggested that the majority of gas within the CMZ is not bound by self-gravity, rather it is bound by the potential produced from the embedded bulge stars. Therefore, despite the gas being very dense, it will not gravitationally collapse to form stars as it would in the Galactic disc, and the CMZ is therefore at a star formation minimum. This idea has been quantified further by \citet{krumholz_2015} and \citet{krumholz_2016}, who predict that significant star formation should take place once the gas becomes self-gravitating. To investigate this, we examine the star formation rates on parsec scales (rather than global scales) within molecular clouds and star formation regions which are believed to be bound by self-gravity.  

\section{The gas mass and embedded stellar populations within individual clouds}\label{embedded}

\subsection{Determining the gas and embedded stellar masses from infrared observations}\label{determing_embedded}

It is clear from Figure\,\ref{sfr_maps} that the luminosity in the Galactic Centre varies significantly over scales as little as a few parsecs, implying that the instantaneously measured local star formation rate and efficiency also vary over similar scales. However, as the individual clouds only harbour specific stages of star formation, we can not apply the SFR-luminosity relations, as these require continuous star formation over $\gg$\,5\,Myr \citep{kruijssen_2014}. In this section we therefore propose an alternative method to measure the embedded stellar masses, which will be used in the following sections to estimate the star formation rate. 

Firstly, the individual clouds which are believed to be bound by self-gravity are identified, and their boundaries are determined. Next, we measure the enclosed infrared bolometric luminosity. This is used to estimated the mass of the most massive star, from which the total embedded population can be extrapolated using a stellar initial mass function. We choose to limit the sources to those within the region $0.18<l<$0.76$^{\circ}$, $-0.12<b<$0.13$^{\circ}$, as this region has both significant cool and warm gas emission, and is known to contain both quiescent (the so called ``dust-ridge'') and actively star-forming regions (e.g. Sgr B2), whilst not suffering significant line-of-sight confusion from prominent sources within the Galactic Centre (e.g. Sgr A$^{*}$). Despite being limited to this ``simple'' region of the Galactic Centre, a certain level of ambiguity is present when identifying the extent for the sources. The interstellar medium is intrinsically hierarchical and the three-dimensional structure of the gas within the Galactic Centre is complex (e.g. \citealp{rathborne_2015, walker_2015, henshaw_2016}; see section\,\ref{boundaries}). It is therefore difficult to impose physically meaningful cloud boundaries in the same way one can separate individual stars. 

We define the boundaries using various warm and cool column density contours (see Figure\,\ref{sfr_maps}), which have be chosen by-eye to best separate different sources. The column density limits, radii and enclosed gas masses for each source is displayed in Table\,\ref{masses_table}. These masses are within a factor of two to those presented by \citet{immer_2012a}, \citet{longmore_2012}, \citet{walker_2015} and \citet{federrath_2016}. The moderate difference is a result of our higher column density boundaries. We investigate the effect of changing the source boundaries in section\,\ref{Uncertainties}, and show that this does not affect the results of this work. 

To estimate the embedded population within each cloud, we first assume that the total infrared luminosity represents the bolometric luminosity from a single massive embedded star. This is a reasonable assumption since the most massive star should dominate the luminosity of a simple stellar population (M\,$\propto$\,L$^{x}$, where ${x}\,\sim\,1-3.5$; e.g. \citealp{mould_1982, salaris_2005}). To estimate the mass of this embedded object (M$_\mathrm{*, max}$), we adopt the bolometric luminosity-mass conversions presented by \citet{davies_2011}. The total embedded population mass (M$_\mathrm{*, tot}$), is extrapolated by solving the following two equations: 
\begin{equation}
1 = \int_{M_\mathrm{*, max}}^{\infty} m ^{-\alpha}\,dm,
\end{equation}
where $\alpha$ = 2.3, and,
\begin{equation}
M_\mathrm{*, tot} = \int_{0.001\,M_\odot}^{\infty} m ^{1 -\alpha}\,dm,
\end{equation}
where $\alpha$\,=\,0.3 for 0.001\,$<$\,$m/$\,\sol\,$<$\,0.08, $\alpha$\,=\,1.3 for 0.08\,$<$\,$m/$\,\sol\,$<$\,0.5, and $\alpha$\,=\,2.3 for $m$\,$>$\,0.5\,\sol\ \citep{kroupa_2001}. Figure\,\ref{rgb} presents a red-green-blue map of the region containing the sources, where the quiescent clouds are in red and (proto-)clusters in blue, over which the embedded and gas masses are labeled. The calculated masses are summarised in Table\,\ref{masses_table}, which are used to determine star formation rates and efficiencies in Section\,\ref{sec:SFE}. 

\subsection{Determining the embedded stellar masses from additional observations}\label{lit_masses}

In addition to measurements from infrared observations, the embedded stellar mass can be inferred from cm/mm wavelengths observations. \citet{walker_2015} have determined the embedded stellar mass within the Sgr B2 region which encompasses the Sgr B2 ``main'', ``north'' and ``south'' \HII\ region complexes. These authors estimate the mass of high-mass stars embedded within the ultra compact \HII\ regions within these complexes (UC\HII) from their 1.3\,cm continuum emission (\citealp{gaume_1995} identified $\sim$\,40 regions). Given that the cm-observations are only sensitive to high-mass stars, the full population is extrapolated using a Kroupa IMF. \citet{walker_2015} estimate a total embedded stellar population mass of $\sim$\,3500\,\sol. \citet{belloche_2013} determined a similar mass of $\sim$\,3900\,\sol when using this same method, with the data of \citet{gaume_1995}. More recently, \citet{schmiedeke_2016} have complied all the available cm-continuum data from the literature (\citealp{mehringer_1993, gaume_1995, depree_1998}; $\sim$~70 regions identified) and follow the above method to determine the total embedded stellar population mass. These authors estimate a mass of $\sim\,3.3\,\times\,10^4\,$\sol\ (shown in parentheses in Table\,\ref{masses_table}), which is an order of magnitude larger than the previously derived values.

To determine the embedded stellar masses within G0.6-0.005 (henceforth G0.6) and Sgr B1, we follow the method used by \citet{walker_2015}, \citet{belloche_2013}, and \citet{schmiedeke_2016}, and the spectral classifications of the UC\HII\ regions as determined by \citet{mehringer_1992}. To convert these into masses, we use the spectroscopic masses of zero age main sequence stars given by \citet{vacca_1996}. We obtain the total mass using a Kroupa IMF. We find that the total embedded masses within G0.6 and Sgr B1 are 3300\,\sol\ and 7200\,\sol, respectively. 

Along with the above VLA observations, we use {\it WMAP} observations to calculate the stellar mass within Sgr B1.\footnote{{\it WMAP} observations can only be used for Sgr B1, as this is the only source identified by \citet{lee_2012}.} {\it WMAP} data has better absolute flux precision, suffers less from spatial filtering and covers a larger frequency range than the VLA observations. From this data an accurate spectral index of the emission can be calculated allowing accurate determination of the relative contributions from free-free, non-thermal and spinning dust emission. From the WMAP source catalogue presented by \citet{lee_2012}, we find that Sgr B1 has an inferred ionising flux of Q\,=\,$0.5\pm\,0.19\,\times\,10^{53}$\,s$^{-1}$. \citet{murray_2010} showed that the ionising flux per stellar mass averaged over the initial mass function is $<{\rm q}>/<{\rm m_{*}}>\,=\,6.3\,\times\,10^{46}$\,\sol$^{-1}$s$^{-1}$. Using this, we find that the total embedded stellar mass within Sgr B1 is $\sim$\,8000\,\sol. 
  
\subsection{Uncertainties}\label{Uncertainties}  

This section includes a discussion of the uncertainties present when estimating the embedded young stellar mass and gas mass.

\subsubsection{Source boundaries}\label{boundaries}

\begin{figure}
\centering
\includegraphics[trim = 4mm 0mm 0mm 0mm, clip,angle=0,width=0.95\columnwidth]{\dir 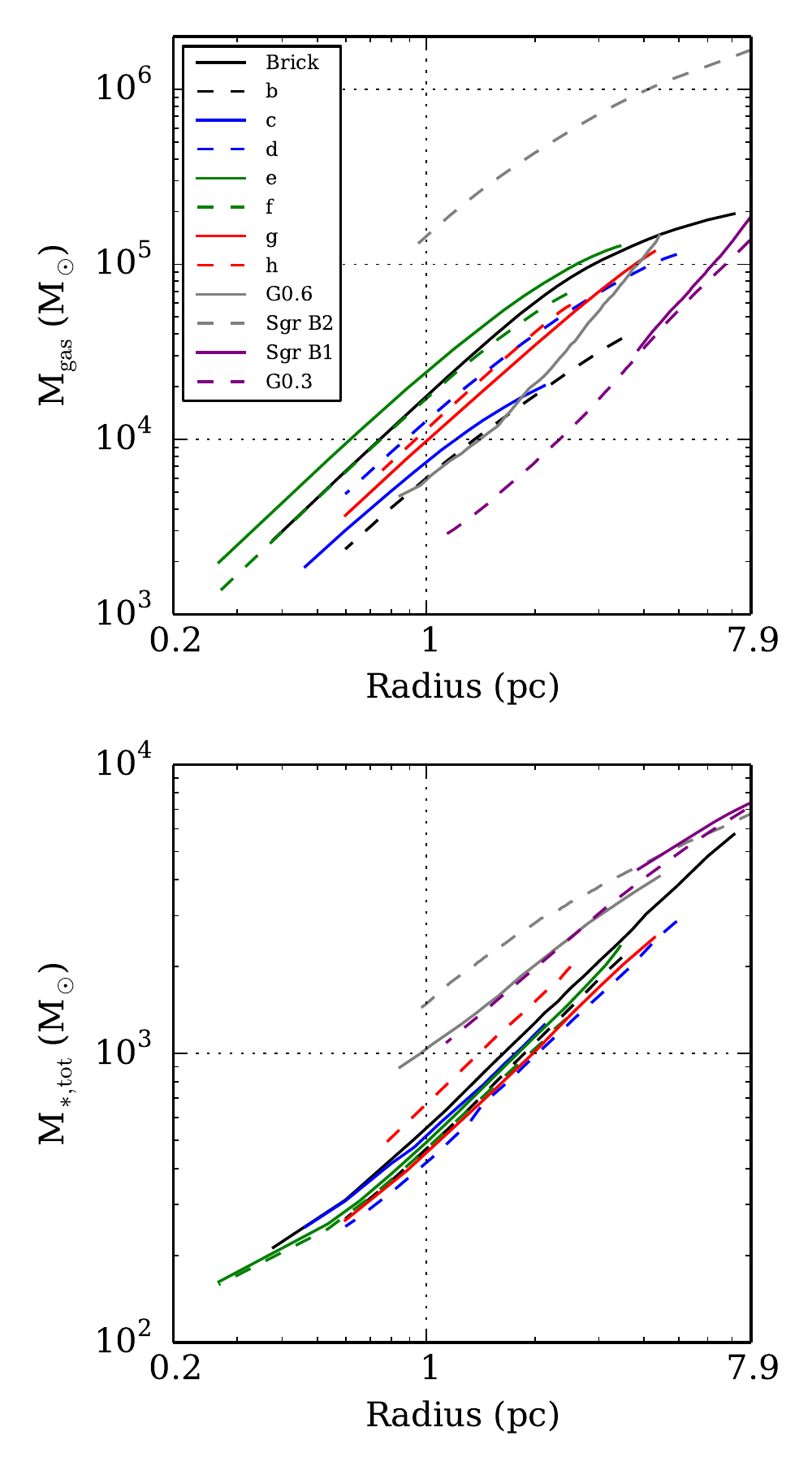}
\caption{Plots of radius against the total mass of the gas (upper panel) and embedded young stellar population (lower panel). The radial profiles of the CMZ sources are shown by the coloured solid and dashed lines (see the legend in the upper panel). } 
\label{radius_vs}
\end{figure}  

In section\,\ref{determing_embedded} we attempted to determine the boundaries of sources within the region $0.18<l<$0.76$^{\circ}$, $-0.12<b<$0.13$^{\circ}$, using several column density contours. This is not trivial as the interstellar medium is intrinsically fractal and hierarchically structured. A similar difficulty in defining the sources was noted by \citet{walker_2015, walker_2016}, who showed that the properties of sources vary depending on the choice of boundary. To test the effect of changing the boundaries, Figure\,\ref{radius_vs} shows how the enclosed gas and embedded stellar masses vary as a function of radius (assuming a spherical geometry). Varying the radius of each cloud by approximately $\pm$\,30\,per cent gives on average a difference of M$_\mathrm{gas}$$^{+90\%}_{-40\%}$, and M$_\mathrm{*, tot}$$^{+50\%}_{-20\%}$. 

We find the Sgr B2 region is particularly difficult to define with a single column density contour, as it is thought to contain both dense gas and actively star forming regions \citep[also see][]{schmiedeke_2016}. We choose to define the Sgr B2 region by a cool component column density of 7.5\,$\times$10$^{23}$\,\,cm$^{-2}$, as this contour separates it from neighbouring sources (see section\,\ref{determing_embedded}). However, this contour does not contain the warm luminosity component to the south-west ($l\approx0.64^{\circ}$, $b\approx\,$-$0.8^{\circ}$) or the extended cool envelope to the north ($l\approx0.6^{\circ}$, $b\approx0.0^{\circ}$), which have both been previously attributed to Sgr B2 (e.g. \citealp{yusef_2009}). Increasing the column density limit to include these would increase the gas and embedded stellar masses by factors of two to three.

Changing the source boundaries varies both the gas mass and embedded young stellar mass on average by a factor of two, however the gradient of both parameters as a function of radius are similar. As the star formation efficiencies per free-fall time which are calculated in the later sections of this work (section\,\ref{sec:SFE}), are essentially ratios of the gas mass and embedded young stellar mass, these are are not sensitive to the choice boundary over a few parsecs.

\subsubsection{Field Star contribution}

The sample of clouds within $0.18<l<$0.76$^{\circ}$, $-0.12<b<$0.13$^{\circ}$ are thought to reside $\sim$\,100\,pc from the centre of the galaxy. There is high number density of old population bulge stars within this region, which may contribute to the measured infrared luminosity and cause an overestimation of mass of the embedded stellar population. To investigate this, we use the Besan\c{c}on model \citep{robin_2003}. We use the same catalogue simulation from section\,\ref{totalsfr}, with a distance range of 8.3\,$-$\,8.5\,kpc and a step of 50\,pc, towards the coordinates of the clouds within the $0.18<l<$0.76$^{\circ}$, $-0.12<b<$0.13$^{\circ}$ region. This gives an average bolometric luminosity density of 28\,$\pm$\,5\,L$_\odot$\,pc$^{-3}$ ($\sim$15\,stars\,pc$^{-3}$). Therefore, we find that the old stellar population stars will contribute to $<$\,1\,per cent of the total bolometric luminosity measured for each cloud. 

\subsubsection{Background contribution}

There is a smoothly varying infrared component to the diffuse Galactic emission along the line-of-sight to the CMZ, which contributes to the luminosity. This has been subtracted for the emission at wavelengths $>$70\micron\ using the method presented in \citet{battersby_2011}, which causes an average decrease of $\sim$20\,per cent in the measured bolometric luminosity at each pixel. There may be some residual diffuse emission that varies on small scales associated with the CMZ itself, which has not been removed when using the background subtraction routine and could contaminate the bolometric luminosity measurements for each cloud. We, however, do not expect this to be a significant effect.

We have not taken into account background subtraction for the line-of-sight emission at wavelengths $<$70\micron, which could lead to an overestimation of the luminosity. This is expected to be more of a problem for the sources which have less emission at these wavelength, such as those in the dust-ridge. We estimate the magnitude of this to be on the order of the $>$70\micron\ Galactic diffuse emission: $\sim$20\,per cent. On the other hand, we do not believe this will be a significant effect to the brighter star-forming clouds (e.g. Sgr B2), as their luminosities are clearly dominated by the embedded young stellar population.

\subsubsection{Luminosity from external heating}

The dust ridge clouds are thought to be externally heated \citep{longmore_2013, longmore_2013a, ott_2014, rathborne_2014a}, therefore we consider that some fraction of their luminosity may be produced by the heating from bright nearby sources. As an example, we estimate how much luminosity would be produced by the heating from the two brightest clusters within the Galactic Centre, the Arches and Quintuplet clusters. Using the three-dimensional structure of the Galactic Centre determined by \citet[][see section\,\ref{sec:SFE}]{kruijssen_2015}, we estimate that the ``Brick'' is the closest cloud, residing at a distance of $\sim$\,25\,pc from these clusters, hence use this as the example subject. \citet{figer_1999, figer_2002} estimate that the luminosity within the Arches and Quintuplet clusters is $10^{7.8}$ and $10^{7.5}$\,\Lsol, respectively. Assuming that the luminosity is isotropically radiated from each cluster, and is completely absorbed and re-emitted by the ``Brick'' (assuming a circular geometry with a radius of $\sim$\,3.1\,pc), the luminosity contribution from the Arches and Quintuplet clusters is $\sim\,2\,\times\,10^{5}$\,\Lsol\ and $10^{5}$\,\Lsol, respectively. This is $\sim$\,30 per cent of the total luminosity of the ``Brick''. We suggest that this is an upper limit to the affect of external radiation on the measured luminosity of the sources considered here.

\subsubsection{Accretion luminosity}

We consider that some fraction of the bolometric luminosity from these clouds may be caused by the accretion of material onto the embedded stars. It is thought that the accretion luminosity for low-mass young stars can be around an order of magnitude higher than the intrinsic stellar luminosity, whereas for high-mass stars the stellar luminosity dominates over the accretion luminosity for all reasonable accretion rates ($>$\,10\,\sol; e.g. \citealp{hosokawa_2009}). In this work we assume that the infrared luminosity from each of the embedded stellar populations is dominated by high-mass stars, so assume the contribution of the accretion luminosity should be insignificant.  

%\vspace{20cm}

\subsubsection{Variation in the embedded stellar population}\label{IMF}

We calculate the total embedded stellar population mass within each source in the Galactic Centre by extrapolating from a high mass sample of the population. The three caveats which may affect our measurement of the embedded stellar population mass are the choice of the initial mass function, the sampling of the initial mass function and the upper mass limit of the initial mass function (i.e. the maximum stellar mass of the population), and are discussed below.

Throughout this work we estimate the embedded stellar population assuming a Kroupa IMF \citep{kroupa_2001}. However, many IMFs are used in the literature, which despite years of major scrutiny, share broadly similar properties: a power-law at the high-mass end with a slope of roughly -2.3, and a turnover (Kroupa) or Gaussian \citep{chabrier_2003} low-mass end. This shape turns out to be universal, from the solar neighbourhood \citep{bastian_2010} to the CMZ (in the Arches cluster; \citealp{habibi_2013}). However, we note that it has been suggested that the nuclear star cluster (i.e. central parsec around Sgr A$^*$) has a top-heavy initial mass function \citep{bartko_2010}, and there is some indirect evidence for a bottom-heavy IMF in early-type galaxies, which harbour a similar environment to the CMZ \citep{conroy_2012, cappellari_2012}.

When using IMFs it is important to consider their stochastic nature, particularly for stellar populations with low number statistics - i.e. in regions where star formation has recently begun. When stars form they stochastically populate the IMF, such that each star has finite probability of having any mass between a given mass range. More massive stellar populations will in general have more stars, and hence will have a higher chance of being fully populated \citep[e.g. ][]{gilmore_2001, fumagalli_2011}. Populations with masses of $\sim$\,10$^{4}$\,\sol\ should fully sample the initial mass function \citep{bruzual_2003}. In this work we estimate embedded stellar masses significantly less than this, for example dust-ridge clouds have M$_\mathrm{*, tot}$\,$\sim$\,10$^3$\,\sol ($\sim$\,400\,-\,1000\,stars). The initial mass function for these sources will, therefore, most likely not be fully sampled, and stochastic effects may be significant. \citet{elmegreen_1999} have shown that low number sampling (1000\,stars) of an IMF could cause the power-law slope to vary by $\pm$\,0.1. This effect is inherently ``random'' and is therefore difficult to quantify for each source. Nevertheless, changing a single power-law slope of $\alpha\,=\,2.3$ by $\pm$\,0.1, would vary the total embedded mass by a factor of two to three.

The final caveat in calculating the total embedded stellar population mass for each source is that the bolometric luminosity is produced solely by the most massive star in the population. There may be, however, a non-negligible contribution by the second, and progressively lower massive stars to the measured bolometric luminosity, which could cause an overestimation of the most massive star in the population. To investigate how this affects the total mass of the embedded population, we plot the bolometric luminosity as a function of total embedded stellar population mass, with the assumption that all the luminosity is produced by the most massive star. On the same axis we plot the bolometric luminosity as a function of total embedded stellar population mass produced from the synthetic stellar population model {\sc starburst99} (see Figure\,\ref{mass_lum}).\footnote{\url{http://www.stsci.edu/science/starburst99/docs/default.htm}} The model input parameters are an instantaneous star formation burst populating a Kroupa IMF with total cluster masses ranging from 1000 to 100,000\,\sol. We note that, the {\sc starburst99} is unable to produce stellar populations with masses below 1000\,\sol. We find that the {\sc starburst99} mass-luminosity relation has a power-law slope which is shallower than if we assume all the luminosity is produced from the most massive star, with the intersection between the two relations at $\sim$\,5000\,\sol. As the star-forming sources G0.6, Sgr B1 and G0.3-0.056 (henceforth G0.3) have masses close to this value, these should not be significantly affected by this uncertainty. However, the dust ridge clouds have embedded masses much lower than this. The {\sc starburst99} modelling shows that stochastic sampling may lead to an overestimation of the total embedded stellar mass by up to factor of 3 (see Appendix\,\ref{Appendix B}).

In summary, we estimate that the main sources of uncertainty on the embedded population within the Galactic Centre clouds are the stochastic nature of the IMF and the form of the adopted IMF. This leads to an uncertainty in the embedded stellar mass estimate of at least a factor of two.

\subsubsection{Saturation}

Several of the {\it Spitzer} and {\it Herschel} maps used in this work contain saturated pixels, which we treat these as having the maximum observed value within each map. This was not considered a problem when determining the global star formation rates as these pixels did not significantly contribute to the total luminosities. However, this is not the case for the individual clouds, where the majority of saturated pixels are located. We find that 8, 5, 40,\footnote{We highlight that the high fraction of saturated pixels in the bolometric luminosity maps towards the Sgr B1 region is a result of saturation in the {\it Spitzer} 24\,\micron\ map.} and 3\,per cent of the pixels within Sgr B2, G0.6, Sgr B1, and G0.3, respectively, are saturated. The luminosity towards these saturated pixels is considered a lower limit, and therefore contributes to an underestimation of the measured embedded stellar mass of each source. We expect that this will cause a more severe underestimation of the embedded stellar mass when a high concentration of the embedded sources are within the saturated pixels. Despite Sgr B1 having the majority of saturated pixels, we find only eight \HII\ regions towards these pixels \citep{mehringer_1992}, whereas within the saturated pixels towards Sgr B2 contain more than sixty \HII\ regions \citep{schmiedeke_2016}. We therefore expect pixel saturation to cause a significant underestimation of the embedded stellar mass towards the Sgr B2 region.

\subsubsection{Summary of embedded stellar population mass uncertainties}

To summarise, several of the uncertainties discussed above can significantly affect the estimate of the embedded stellar population masses for all the sources. Some are only applicable to either the quiescent clouds (high gas mass/low embedded stellar mass), or the star-forming clouds (low gas mass/high embedded stellar mass). The uncertainties which affect all the sources are the choice of arbitrary boundary ($\pm$50\,per cent) and the contribution from accretion luminosity (negligible). The uncertainties which primarily affect the quiescent clouds are the contribution of CMZ diffuse background luminosity and the stochasticity in the IMF, as their low masses do not enable full/significant sampling of the IMF. As the diffuse background luminosity will cause the stellar mass to be overestimated, we conclude that the embedded stellar mass estimates for the quiescent clouds (``Brick'', ``b'', ``c'', ``d'', ``e'' and ``f'') should be considered as upper limits. On the other hand, the main uncertainty on the bolometric luminosity, hence the embedded stellar mass, which affects the star-forming clouds is pixel saturation, yet it is difficult to quantify the magnitude of the uncertainty this induces. To approximate severity of this, we compare the embedded stellar masses determined from the infrared bolometric luminosity and cm-continuum observations (section\,\ref{lit_masses} and Table\,\ref{masses_table}). We find that the discrepancy in embedded stellar mass determined from these observations for G0.6 and Sgr B1 is small ($20-30$\,per cent). Therefore, given the other uncertainties on the embedded stellar mass discussed in this section, we conservatively estimate that the embedded stellar masses determined for these two sources are reliable to within a factor of $2-3$. As no previous measurements of embedded stellar mass toward G0.3 are available, we suggest that this is also reliable to within a factor of $2-3$. We find that the embedded stellar mass determined from infrared measurements of Sgr B2 is an order of magnitude below the value determined by \citet{schmiedeke_2016}, which we expect is due to the high concentration of UC\ion{H}{II} regions towards the saturated pixels within this source. We suggest that this has caused a significant underestimation of the embedded stellar mass within the Sgr B2, hence this infrared measurement should be considered a lower limit.

\begin{figure*}
\centering
\includegraphics[trim = 0mm 0mm 0mm 0mm, clip,angle=0,width=1\textwidth]{\dir 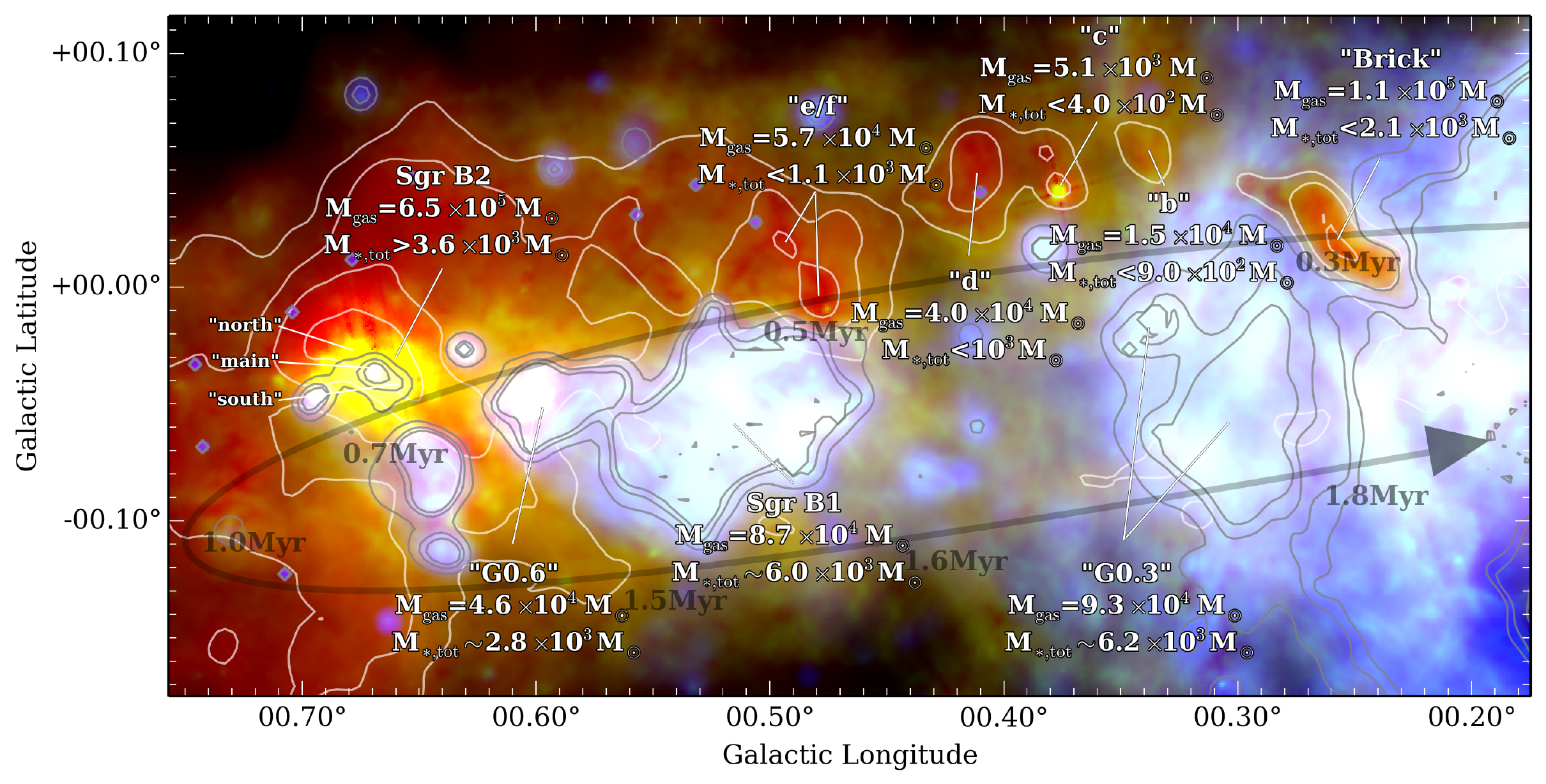}
\caption{Three colour zoom-in of the ``dust-ridge'', shown in blue and red are the warm and cool component luminosities, and in green is the 70\micron \ emission. Over-plotted are contours identical to Figure\,\ref{sfr_maps}. Labeled are the sources with their total gas and embedded masses determined from infrared measurements (see Table\,\ref{masses_table} from embedded masses determined via additional methods). The transparent curved arrow represents the path of the orbital model of \citet{kruijssen_2015}, with labels of time since pericentre passage according to this model.} 
\label{rgb}
\end{figure*}

\section{Deriving the star formation efficiency and star formation rates}\label{sec:SFE}

Having measured the embedded stellar populations and gas masses for each of the regions we now seek to determine the efficiency with which gas is converted into stars in the Galactic Centre. We are interested in deriving two numbers in particular: (i) the ratio of a cloud's gas mass, M$_{\rm gas}$, to the mass in stars, M$_\mathrm{*, tot}$, which is defined as the star formation efficiency,  SFE\,$\equiv$\,M$_\mathrm{*, tot}$/(M$_\mathrm{gas}$ + M$_\mathrm{*, tot}$); (ii) the fraction of the cloud's gas mass which is converted into stars per free-fall time, $\epsilon_\mathrm{ff}\,=\,\mathrm{SFE}\times\,(t_\mathrm{ff}$\,/\,$t_0$), where $t_0$ is the time since the onset of star formation. 

Table\,\ref{masses_table} summarises the properties of gas and young stars in several of the most massive and dense clouds and (proto-)clusters within the Galaxy. Figure\,\ref{rgb} shows the cool molecular gas (red) and hot (ionised) gas (blue and green) towards the region containing these sources of interest; $0.18<l<$0.76$^{\circ}$, $-0.12<b<$0.13$^{\circ}$. Over-plotted are contours of cool and warm gas column densities in white and grey, respectively (see section\,\ref{method}). Labeled on this plot are the sources and their respective gas and embedded stellar population masses within the radius defined in Table\,\ref{masses_table}. 

From the observed M$_{\rm gas}$ and M$_\mathrm{*, tot}$ it is straightforward to derive and compare the SFE for each region. We know that all the regions lie in the same environment, hence we can remove a major source of uncertainty that has hampered previous comparisons of  SFEs for sources that may have formed in (potentially very) different environments. As shown in Table\,\ref{masses_table}, we find that the ``dust-ridge" clouds have SFE\,$<4$\,per cent (upper limits), and the star-forming sources have SFE\,$\sim$\,7\,per cent. Taking these numbers at face value it appears that, despite these regions containing some of the most vigorous star formation activity in the Galaxy, only a small fraction of their total gas mass has so far been converted to stars. %{\bf Surprisingly, the values of the SFE measured here are similar to those measured in clouds in the Galactic disc. For example, Table 3 in \citet{federrath_2013} shows that the SFE is typically of the order of $0-5$\,per cent for most clouds, and only in the very dense star-forming cores does the SFE reach values of several tens of per cent.}

\subsection{Deriving the star formation efficiency per free-fall time assuming a model of tidally triggered star formation}
\label{sub:SFE_triggered}

Other than assuming all the sources lie at the same distance and in the same general environment, all of the analysis until this point has implicitly treated all regions as independent. We now attempt to infer the possible relationship between gas clouds and (proto-)clusters by interpreting pertinent observational facts. 

Firstly, observations of dense gas molecular line tracers (e.g. HNCO, N$_2$H$^+$) towards the region from Figure\,\ref{rgb}, clearly show the quiescent clouds (red) and (proto-)clusters (blue) are all linked along a coherent velocity structure, or ``stream", in position-position-velocity (PPV) space (e.g. \citealp{kruijssen_2015, henshaw_2016}). The quiescent clouds found towards the north of this region lie at one end of the PPV stream. These clouds show very little sign of star formation, despite their similarly large gas masses and small radii \citep[e.g.][]{lis_1999, immer_2012, longmore_2013a, walker_2015, ginsburg_2015}. Following the stream to higher galactic longitudes and velocities, and lower latitudes lies the mini-starburst complex Sgr B2 (e.g. \citealp{bally_1988, hasegawa_1994, sato_2000}). Continuing on the stream from Sgr B2 to lower longitudes, latitudes and velocities, the \HII\ region complexes G0.6, Sgr B1 and G0.3 are found (e.g. \citealp{mehringer_1992, lang_2010}). The first assumption we make is that as the quiescent clouds and (proto-)clusters are all part of the same stream.

Secondly, the quiescent gas clouds all have very similar masses and radii, and are close to virial equilibrium and therefore likely to collapse \citep{walker_2015}. Thirdly, the gas column density probability distribution functions of the quiescent gas clouds are indicative of imminent star formation (\citealp{rathborne_2014}; Battersby in prep.), and the radial distributions of the mass surface densities suggest that collapse of these clouds could produce Arches or Quintuplet-like clusters in the future \citep{walker_2016}. Lastly, there is a general progression of star formation along the stream from quiescent gas clouds to (proto-)clusters.

If the clouds and (proto-)clusters within the $0.18<l<$0.76$^{\circ}$, $-0.12<b<$0.13$^{\circ}$ region do represent an evolutionary sequence, one has an estimation of the initial conditions for star formation within the Galactic Centre. With a measure of SFE from section\,\ref{sec:SFE} and $t_{\rm ff}$, the only thing needed to derive $\epsilon_\mathrm{ff}$ is an absolute timescale linking the clouds and (proto-)clusters.

\citet{kruijssen_2015} have developed a dynamical orbital solution to interpret the PPV structure of the molecular emission throughout the central $\sim\,$200\,pc of the Galaxy, from which ``Streams 2 and 3'' have been over-plotted in Figure\,\ref{rgb}. The focal point of this open, elliptical orbit coincides with the position of the supermassive black hole at the centre of the Galaxy, Sgr A$^*$, and all the sources within the $0.18<l<$0.76$^{\circ}$, $-0.12<b<$0.13$^{\circ}$ region are downstream past pericentre passage on this orbit.

The observed star formation activity increases with time past pericentre (shown as the increasing spatial extent of the hot gas component in Figure\,\ref{rgb}). In the scenario presented by \citet{longmore_2013a}, \citet{kruijssen_2015}, and \citet{longmore_2016}, gas clouds will experience strong tidal forces close to pericentre passage, which will compress the gas along the vertical direction. This adds turbulent energy into the gas which can be quickly radiated away by shocks due to the high density. In the model, this implies the clouds become self-gravitating and allows them to initiate gravitational collapse, eventually triggering star formation. Further downstream, as star formation continues, feedback from massive embedded stars begins to blow cavities in the surrounding environment, and eventually cause the dissipation of the host cloud (Barnes et al. in prep). 

Assuming that star formation within each cloud was triggered at pericentre passage, we can calculate the time-averaged star formation rate as SFR\,(\sol\,yr$^{-1}$)\,=\,M$_\mathrm{*, tot}$(\sol)\,$/t _\mathrm{p,last}$\,(yr), where $t _\mathrm{p,last}$ is the time since the last pericentre passage. In Figure\,\ref{rgb} we label the time since pericentre passage for the orbital model presented by \citet{kruijssen_2015}, where the pericentre of the orbit is located just upstream from the ``Brick''. We find the star formation rates of these clouds are in the range $0.001-0.045$\,\solyr. The total star formation rate within these clouds sums to $0.03-0.071$\,\solyr. These clouds, therefore, contribute a quarter to three-quarters of the total star formation within the CMZ ($|l|$$<$1$^{\circ}$, $|b|<$0.5$^{\circ}$; $\sim$\,0.09\,\solyr). The values for the individual clouds are presented in Table\,\ref{SFR_table}, and are in agreement with those determined through independent methods by \citet[][also see Table\,\ref{SFR_table}]{kauffmann_2016a} and Lu et al. (in prep).

The calculated star formation rates are used to determine the fraction of the cloud's mass which is converted into stars per free-fall time. This can be described as $\epsilon_\mathrm{ff}\,=\,\mathrm{SFE}\times\,(t_\mathrm{ff}$\,/\,$t_\mathrm{p, last}$). We find star formation efficiencies per free-fall time in the range of $1 - 5$\,per cent,\footnote{Note that these values change by less than a factor of two when using typical timescales for star formation in young massive clusters (e.g. the free-fall time; see the review by \citealp{longmore_2014}), rather than the timescales from the \citet{kruijssen_2015} orbital model.} which are again listed in Table 5.

\subsection{Comparison to theoretical models}\label{Comparison to theoretic models}

We now compare these measurements of the star formation rate to the predictions of star formation theories within the literature. As discussed in section\,\ref{models}, we limit our comparison to the column density threshold relation and volumetric models for star formation (see reviews by \citealp{federrath_2012} and \citealp{padoan_2014}). The column density threshold relation predicts the star formation rate solely from the dense gas mass \citep{lada_2010, lada_2012}. On the other hand, the volumetric models predict the dimensionless star formation efficiency per free-fall time, $\epsilon_\mathrm{ff}$, from the physical global properties of the cloud: the sonic Mach number, Alfv{\'e}nic Mach number, virial parameter, and the turbulence driving parameter. \citet{padoan_2014} show that a comparison between the predictions from these relations/models and observations is not trivial. Scatter of more than an order of magnitude exists in the observed star formation rate for a given gas mass, due to the range of environments, evolutionary stages, and spatial scales probed by observations of different clouds. 

In section\,\ref{sub:SFE_triggered} we discussed several means by which we hope to overcome several of these limitations: i) directly determining the gas mass and embedded stellar population mass within several sources in the same environment and at the same spatial resolution, ii) using the \citet{kruijssen_2015} orbital model to estimate the absolute time which has passed since star formation within each source was plausibly triggered, and iii) limiting our study to sources within the extreme environment of the Galactic Centre, which may be representative of the environment in which the majority of stars have formed \citep{kruijssen_2013}. 

%{\bf Only recently has observations and star formation theory been compared within a sample of galactic centre clouds (e.g. \citealp{federrath_2016, kauffmann_2016a}).}

\vspace{0.5cm}

\citet{lada_2010, lada_2012} propose that the gas above a column density of $6.7\,\times\,10^{21}\,$cm$^{-2}$ has a universal depletion timescale of $\sim$\,20\,Myr ($\sim$\,$50$\,$t_\mathrm{ff}$). They propose that the star formation rate can be predicted using Equation\,\ref{lada}. We find that the constraints on the star formation rate within the dust-ridge clouds ``d'', ``e'', and ``f'' are too poor to test the \citet{lada_2010, lada_2012} predictions with any significance. \citet{kauffmann_2016a} have inferred an upper limit of the star formation rate within the ``Brick'' of $<$\,0.0008\,\solyr, from the non-detection of the star formation tracers (e.g. \HII\ regions and masers). This is an order of magnitude below the upper limit found here, and is most likely more representative of the true star formation rate within this source. The star formation rate predicted by the column density limit relation for the ``Brick'' is $\sim$\,0.006\solyr, which is significantly higher that measured by \citet{kauffmann_2016a}. The predictions for the star-forming sources Sgr B2, G0.6, Sgr B1 and G0.3 show better agreement with the observed star formation rates.

%The predictions for the star-forming sources Sgr B2, G0.6, Sgr B1 and G0.3 of 0.033, 0.002, 0.004, and 0.005\,\solyr\ respectively, are in good agreement with the observed star formation rates.} %However, whether or not these more evolved clouds have star formation rates consistent with the predictions depends sensitively on the boundaries used to define the outer radius of the cloud and which method is used to determine the star formation rate.
 
Predictions from volumetric star formation models within the Galactic Centre have been previously determined, however their comparison to observations have been limited. \citet{rathborne_2014} and \citet{rathborne_2015} used The Atacama Large Millimeter/submillimeter Array to measure 3\,mm continuum dust emission at high-spatial resolution towards the ``Brick''. These authors measure a column density probability function dispersion of $\sigma_N$\,=\,0.34, and place lower limits on the critical over-density of collapse of $x_{crit}>100$ (see equation\,\ref{pdf}). Although the star formation rate was not calculated, these results are consistent with the an environmentally dependent {\it absolute} density threshold for star formation, which is orders of magnitude higher than that derived for clouds within the disc of the Milky Way. Recently, \citet{federrath_2016} have conducted a further analysis of these continuum data, with the addition of molecular line observations from \citet{rathborne_2015} and dust polarisation observations from \citet{pillai_2015}. These observations were used to determine the dimensionless parameters required for the volumetric star formation models: the three-dimensional turbulent Mach number, the virial parameter including both turbulence and shear, the turbulent magnetic field parameter, and the turbulence driving parameter (see section\,\ref{models}). These authors used the multi-free-fall model of \citet{krumholz_2005}, with the fiducial parameters determined by \citet{federrath_2012}, to predict a star formation efficiency per free-fall time within the brick of $\epsilon_\mathrm{ff}$\,=\,4\,$\pm$\,3\,per cent. This is consistent with our measurement of $\epsilon_\mathrm{ff}\,\sim$\,2\,per cent towards the ``Brick'' (see Table\,\ref{masses_table}). We note, however, that the observed $\epsilon_\mathrm{ff}$ within the ``Brick'' should be considered an upper limit. Taking the embedded stellar population inferred by \citet{kauffmann_2016a} would give $\epsilon_\mathrm{ff}\,\sim$\,1\,per cent, which is only just within the lower uncertainty range of the value predicted by \citet{federrath_2016}. Here we attempt to expand on this comparison by testing all the volumetric star formation models presented by \citet{federrath_2012} against the observed star formation within a number of Galactic centre clouds.

To compare the measured star formation rates to the volumetric models, we first need to determine the initial conditions for star formation for each source. As discussed in section\,\ref{sub:SFE_triggered} we assume that the ``dust-ridge'' clouds should represent the early evolutionary stages of the star-forming sources. The ``Brick'' is the most recent of the ``dust-ridge'' clouds to pass pericentre, hence its properties should best represent the initial conditions for star formation within this region. This cloud has the benefit of being the most well studied of the dust-ridge clouds (i.e. its properties are the most well constrained), and is known to contain little-to-no active star formation (e.g. \citealp{kauffmann_2013, johnston_2014, rathborne_2014, kauffmann_2016a}). 

However, we note that the volumetric models are limited to predicting the $\epsilon_\mathrm{ff}$ over the next free-fall time. In the orbital model of \citet{kruijssen_2015}, the ``Brick'' and Sgr B2 are separated by a time along the orbit that is similar to the free-fall time of the ``Brick'' (see Tables\,\ref{masses_table} and \ref{SFR_table}), and thus the comparison for these regions should be robust. Given that there are several free-fall times between the ``Brick'' and G0.3, using the initial conditions present within the ``Brick'' to predict $\epsilon_\mathrm{ff}$ in the more evolved sources (G0.6, Sgr B1 and G0.3) may not be ideal. Nevertheless, given the relatively small range in gas properties of the ``dust-ridge'' clouds and of the progenitor condensations upstream from pericentre \citep{henshaw_2016e}, the conditions within the ``Brick'' should provide at least approximate predictions for these more evolved sources, which can be compared to the observations. 

The star formation rates determined for the quiescent clouds are not considered for the comparison to the volumetric models, as their embedded masses measured from infrared observations (and hence $\epsilon_\mathrm{ff}$) are considered as upper limits (see section\,\ref{Uncertainties}). We, therefore, limit our analysis to the star-forming sources (Sgr B2, G0.6, Sgr B1, and G0.3), for which we are confident in the measurement of the embedded stellar mass.

\defcitealias{krumholz_2005}{KM05}
\defcitealias{padoan_2011}{PN11}
\defcitealias{hennebelle_2013}{HC13}

\begin{figure*}
\centering
\includegraphics[trim = 0mm 7mm 0mm 3mm, clip,angle=0,width=0.86\textwidth]{\dir 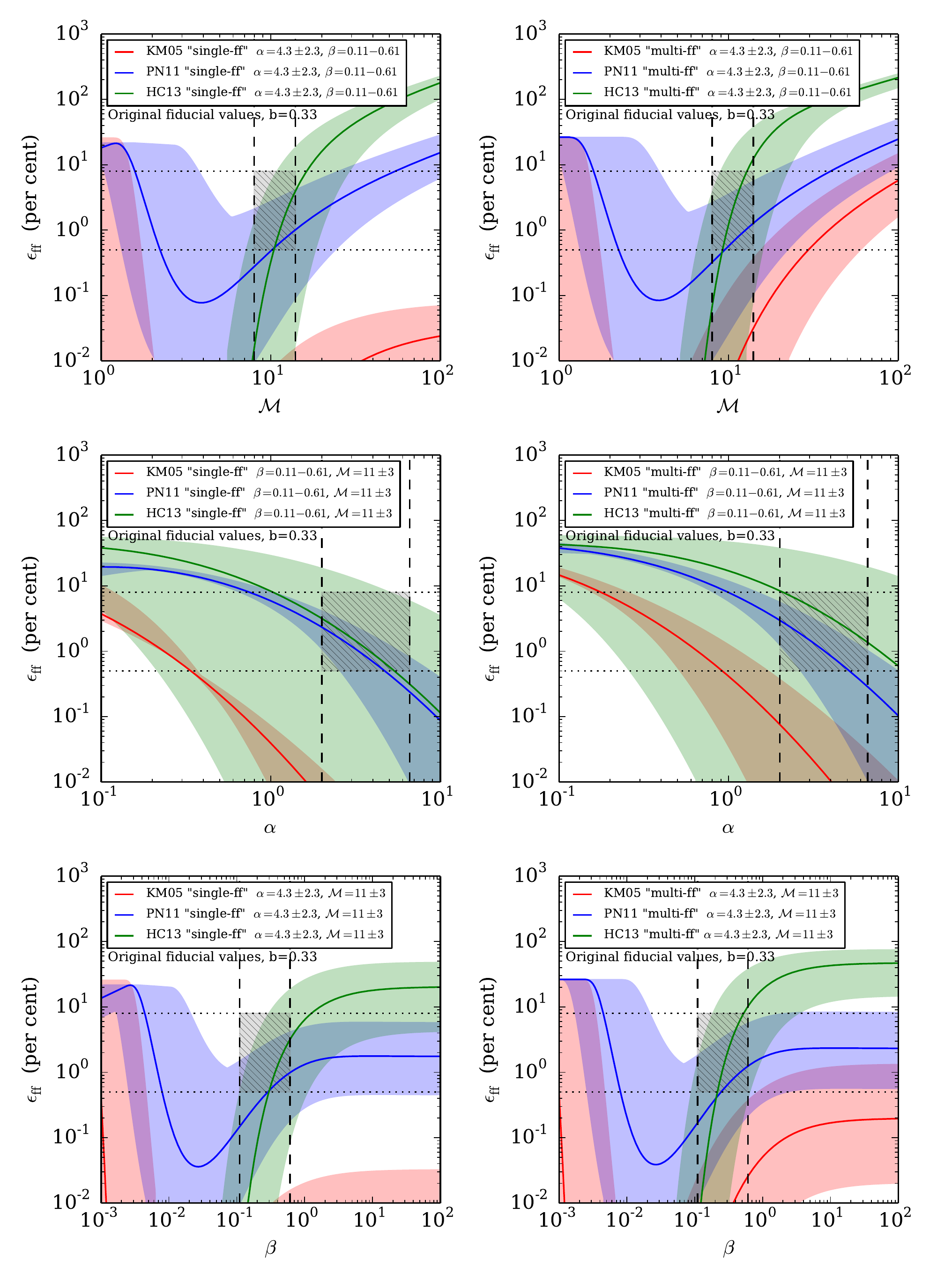}
\vspace{-2.5mm} 
\caption{Plots of the star formation efficiency per free-fall time, $\epsilon_\mathrm{ff}$, as predicted from the single-free-fall (left panels) and multi-free-fall (right panels) models, as a function of the Mach number (upper panels), the virial parameter (middle panels), and the magnetic field strength (lower panels). The purely hydrodynamical scenario is retrieved by setting $\beta\rightarrow\infty$. These are calculated using the fiducial values of $\phi_x\approx1.12$ \citep[KM05]{krumholz_2005}, $\theta\approx0.35$ \citep[PN11]{padoan_2011}, and $y_\mathrm{cut}\approx0.1$ \citep[HC13]{hennebelle_2013}. The coloured lines represent the model predictions. The shaded coloured regions represent the upper and lower limits within the range of our adopted initial conditions (as shown in the legend of each plot). The vertical dashed lines show the result of varying the variable on the x-axis by the range assumed to represent the properties present within the ``Brick'', i.e. the initial conditions for star formation within this region: $\beta = 0.11-0.61$, $\mathcal{M} = 11 \pm 3 $, $\alpha = 4.3 \pm 2.3$ \citep{federrath_2016} and $b = 0.33$ (see text). Here we adopt values of $\epsilon_{core}\approx0.5$ and $\phi_{t}\approx1.91$ for each model \citep{federrath_2012}. The horizontal dotted regions represent the range of $\epsilon_\mathrm{ff}$ for the star-forming sources within the $0.18<l<$0.76$^{\circ}$, $-0.12<b<$0.13$^{\circ}$ region (determined from both infrared and VLA embedded stellar mass estimates, see Table\,\ref{SFR_table})\textsuperscript{\ref{footnote1}}, accounting for the approximate factor of two uncertainty in $\epsilon_\mathrm{ff}$ (i.e. $\epsilon_\mathrm{ff}$ = 0.5 -- 8 per cent; see section\,\ref{Uncertainties}).}
\label{sfr_ff_vs_beta}
\end{figure*}

We use the volumetric models given in the form presented by the \citet{federrath_2012} review (see their Table\,1; also see \citealp{padoan_2014}). This includes the derivation of the log-normal volume density probability distribution function (equation\,\ref{sigma}), and the critical density for collapse and solution to the $\epsilon_\mathrm{ff}$ for each model (equation\,\ref{pdf}). We refer to each model based on its original reference  --  \citet[][KM05]{krumholz_2005}, \citet[][PN11]{padoan_2011}, and \citet[][HC13]{hennebelle_2013} -- and whether the derivation is from the single-free-fall or multi-free-fall form (see section\,\ref{models}). Figure\,\ref{sfr_ff_vs_beta} shows how the predicted star formation efficiency per free-fall time varies as a function of the three-dimensional turbulent sonic Mach number, $\mathcal{M}$, the (turbulence+shear) virial parameter, $\alpha$, and the parameter which describes the strength of the turbulent magnetic field, $\beta$ (the purely hydrodynamical scenario is retrieved by setting $\beta\rightarrow\infty$). Here we use a value of $b$\,=\,0.33 which represents a purely solenoidal turbulent driving mode, as \citet{krumholz_2015} suggest that shear is the typical driving mode of turbulence within the dust-ridge clouds, and within clouds at the centres of other galaxies. The vertical dash lines shows the result of varying either $\mathcal{M} = 11 \pm 3 $, $\alpha = 4.3 \pm 2.3$, and $\beta = 0.11-0.61$,\footnote{The probability distribution function of $\beta$ within the ``Brick'' is asymmetric, with a mean value and standard deviation of $\beta = 0.34\,(0.35)$ \citep{federrath_2016}. Here we use the 16$^{\rm th}$ and 84$^{\rm th}$ percentiles of this distribution ($\beta = 0.11-0.61$; C.~Federrath, private communication).} which are representative of the conditions derived for the ``Brick'' \citep{federrath_2016}. Each model line (shown in colour) is plotted as a function of one of the above three parameters, where the shaded region around each line indicates the uncertainty on the predicted $\epsilon_{\rm ff}$ when varying the two remaining parameters over the ranges specified above. The horizontal dotted lines show the calculated ranges of $\epsilon_\mathrm{ff}$, within star-forming regions Sgr B2, G0.6, Sgr B1, and G0.3.\footnote{As $\epsilon_\mathrm{ff}$ measured from infrared observations for Sgr B2 is a lower limit, henceforth we use only the more accurate $\epsilon_\mathrm{ff}$ determined using the VLA embedded stellar mass \citep{schmiedeke_2016}. \label{footnote1} } Models passing through the shaded region enclosed by the dotted and dashed lines correctly predict the star formation rate per free-fall time in the star-forming clouds, when assuming that their initial conditions were similar to the current properties of the ``Brick''. We find that both single-free-fall and multi-free-fall models of \citetalias{padoan_2011} and \citetalias{hennebelle_2013} enter the shaded region on Figure\,\ref{sfr_ff_vs_beta}.

%\footnote{\label{refnote} For consistency between the models presented by \citet{krumholz_2005}, \citet{padoan_2011} and \citet{hennebelle_2013}, we adopt values of $\epsilon_{core}\approx0.5$ and $\phi_{t}\approx1.91$ \citep[see][]{federrath_2012}.}

%\footnote{{\bf Throughout this work we take the upper limit of $b\,=\,0.22\,\pm\,0.12$ from \citet{federrath_2016}, which represents purely solenoidal turbulence driving. \citet{krumholz_2015} suggest that shear is the typical driving mode of turbulence within the dust-ridge clouds, and within clouds at the centres of other galaxies.}}

%\vspace{20cm}

When testing the volumetric star formation theories, we need to choose the value for several of their free parameters ($\epsilon_{core}$, $\phi_t$, $\phi_x$, $\theta$, and $y_\mathrm{cut}$). In Figure\,\ref{sfr_ff_vs_beta} we adopted the fiducial parameters from the original papers, which have been summarised in Table\,\ref{fiducial values}. \citet{federrath_2012} have constrained these parameters in a different way, by fitting them to magneto-hydrodynamical turbulent box simulations, resulting in the substantially different values, which are also shown in Table\,\ref{fiducial values}.\footnote{We note that the \citet{federrath_2012} determine values of $y_\mathrm{cut}\,>\,1$, however a physical interpretation of $y_\mathrm{cut}$ can only be made for $0\,>\,y_\mathrm{cut}\,>\,1$ (\citetalias{hennebelle_2013}).} 
The result of using the \citet{federrath_2012} parameters is shown in Figure\,\ref{sfe_ff_models-vol-FK12values-b033}, where we find {\it all} of the volumetric models appear to overpredict the star formation rate by factors of several. Recently, \citet{federrath_2016} used the \citet{federrath_2012} fiducial values and a turbulent driving parameter of $b\,=\,0.22$\footnote{The idealised case presented by \citet{federrath_2008}, $b\,=\,0.33$ represents purely isotropic solenoidal turbulence driving. However, lower values of $b$ may be possible when the driving is anisotropic, whereby a particular vortex direction is continuously driven (e.g. by anisotropic shear).} 
 in the \citetalias{krumholz_2005} multi-free-fall model to predict the $\epsilon_\mathrm{ff}$ within the ``Brick''. Figure\,\ref{sfe_ff_models-vol-FK12values-b022} shows the result of using these values for all the volumetric star formation theories. Here we find that both the single-free-fall and multi-free-fall models of \citetalias{krumholz_2005} and \citetalias{padoan_2011} enter the shaded region, and are therefore consistent with the observed star formation rate. 

The analysis in Appendix\,\ref{Appendix C}, shows that varying the turbulent driving parameter in the range determined for the ``Brick'' ($b = 0.10 - 0.34$; \citealp{federrath_2016}), and adopting either the fiducial values from the original papers or those determined by \citet{federrath_2012} gives a range of predicted star formation rate which spans several orders of magnitude (see Figure\,\ref{sfe_ff_models-vol-b}). In view of the above discussion, we point out that the verification or falsification of these star formation theories is fundamentally obstructed by the lack of consensus on the values of these free parameters.

\section{DISCUSSION AND CONCLUSIONS}\label{conclusions} 

In this work, we have utilised infrared {\it Spitzer} and {\it Herschel} observations, with the aim of investigating the lack of star formation within the extreme environment of the Galactic Centre (see \citealp{longmore_2013}). To do this we have determined the star formation rate for the Galactic Centre as a whole (defined as $|l|$$<$1$^{\circ}$ and $|b|<$0.5$^{\circ}$), by using a variety of extra-galactic luminosity-star formation-rate conversions (which trace star formation within the last $\sim$\,5\,Myr). From the 24\micron, 70\micron, and the total infrared bolometric luminosity (determined from two component modified Planck function fits between $5.8 - 500$\,\micron), we find average global star formation rates of $\sim$\,$0.09\pm0.02\,$\solyr. These are comparable to previous measurements made from YSO counting and the free-free emission, which are sensitive to star formation over the last $\sim\,0.1 - 3$\,Myr. The conclusions that can be drawn from this analysis, are discussed below.

\begin{itemize} 
\item[i)] We can rule out that systematic uncertainties in the star formation rate measurements are causing the apparent low star formation rate, unless this uncertainty affects all the methods in the same way, which seems unlikely. 

\item[ii)] The luminosity-SFR relations and YSO counting methods are consistent in the Galactic Centre. As the Galactic Centre is the most extreme environment for which is it possible to resolve individual forming stars and make this measurement, the results here provide confidence that the luminosity-SFR relations reliably trace the star formation rate over kpc scales within similar environments present in starburst galaxies, and high-redshift galaxies.

\item[iii)] The global star formation rate has not changed by more than a factor of two to three from its current rate over the past few Myr. This is quantitatively consistent with models predicting that the star formation rate is episodic \citep{kruijssen_2014a}, with a timescale of $\sim$\,$10-20$\,Myr \citep{krumholz_2015, krumholz_2016}.
\end{itemize} 
In an attempt to better understand the origin of the presently low star formation within the Galactic Centre, we have investigated the properties of several individual clouds and \HII\ region complexes. These are thought to be at differing evolutionary stages and connected along a coherent gas stream within Galactic Centre \citep[e.g.][]{henshaw_2016}. In order to interpret the observations two assumptions are made about how the gas clouds are related, and how star formation proceeds in this environment. 

We make the assumption that the sources are orbiting along a coherent gas stream with known orbital parameters \citep{kruijssen_2015}, and that star formation within these sources is triggered at the pericentre of the orbit (i.e. when compressive tidal forces are strongest; see \citealp{longmore_2013a, kruijssen_2015}). The direct consequence of this scenario is that the regions reside on a common evolutionary timeline, which allows us to derive their star formation timescales and efficiencies. We estimate that the dense molecular clouds remain relatively quiescent for $0.3-0.5$\,Myr after star formation is triggered, as these contain a stringent upper limit of few hundred solar masses of embedded stars (see Table\,\ref{SFR_table}, ``Brick'' to cloud ``f'', i.e. the ``dust-ridge'' clouds). These then transition to very actively star-forming clouds, which contain a few thousand solar masses of stars, within only $0.2-0.4$\,Myr (as seen towards Sgr B2). The feedback from these stars provides sufficient pressure to remove the remaining dense gas over a timescale of $\sim$\,0.9\,Myr (Barnes et al. in prep), which reveals the later stages of star formation (e.g. diffuse \HII\ regions). We note that this division in phases represents the broad brush strokes according to which star formation proceeds in this region, and that the detailed physical picture will be considerably more complex. Nevertheless, this serves as a general model which can be refined in the future.

We take this simple star formation model for the CMZ and determine star formation rates for the sources in the $0.18<l<$0.76$^{\circ}$, $-0.12<b<$0.13$^{\circ}$ region. We find that on average the quiescent ``dust-ridge'' clouds have stringent upper limits of $<$\,0.007\,\solyr, whereas the ``star-forming clouds'', Sgr B2, G0.6, Sgr B1 and G0.3, have star formation rates in the range $\sim$\,$0.002-0.045$\,\solyr. We find that $\sim$\,$1 - 5$\,per cent of a clouds gas mass is converted to stellar mass per free-fall time. 

We use this Galactic Centre gas cloud data to quantitatively test the predictions of different star formation models/relations. We find that the \citet{lada_2010, lada_2012} column density limit relations significantly under predict the observed star formation rate in the quiescent clouds. The predictions for the star-forming sources (Sgr B2 etc), are in better agreement with the observed values. 

As a first comparison to the volumetric models, we take the predictions presented by \citet{federrath_2016}. These authors use the multi-free-fall model of \citetalias{krumholz_2005} with the fiducial values of \citet{federrath_2012} to predict $\epsilon_\mathrm{ff}$\,=\,4\,$\pm$\,3\,per cent within the ``Brick'', which is consistent with the observed value of $\epsilon_\mathrm{ff}\,\sim$\,2\,per cent. 

Expanding on this, we compare our observed star formation rates to all the volumetric relations. Figure\,\ref{sfr_ff_vs_beta} shows that the \citetalias{krumholz_2005} model does not accurately predict the star formation rate for any set of initial conditions, when using the fiducial values from the original models. 

The middle row of Figure\,\ref{sfr_ff_vs_beta} shows that the \citetalias{hennebelle_2013} model is much more sensitive to variations in ${\mathcal M}$ and $\beta$ than the other models, which is a key signature of the \citetalias{hennebelle_2013} theory (and highlights the role of (shock)-turbulence in star-formation). The observational uncertainty in these properties for the ``Brick'', therefore produce a large range of predicted $\epsilon_\mathrm{ff}$ values for a fixed $\alpha$. This makes verifying/falsifying the \citetalias{hennebelle_2013} model predictions more difficult than for the other models. 

The figures in Appendix\,\ref{Appendix C} show the effect of using the fiducial values of \citet{federrath_2012}, and choosing a value for the turbulent driving parameter of 0.33 and 0.22. The former value represents total solenoidal turbulent driving, whilst the latter value is that determined for the ``Brick'' by \citet{federrath_2016}. The $\epsilon_\mathrm{ff}$ typically changes by an order of magnitude when adopting this range of the turbulent driving parameter. This again highlights the importance for self-consistently determining the values of the model free parameters. 

In general, we find better agreement with the multi-free-fall models over the single-free-fall models, which could ultimately reflect their more accurate description of the hierarchical collapse of star-forming regions. In the future, we aim to use the unique laboratory of the Galactic Centre environment to test additional star formation models.

%{\bf The simplistic method we have used to propagate the observational uncertainties into the dimensionless variables predicted by the models 

We find that the most promising range of parameter space for verification or falsification of the models is at the high ${\mathcal M}$, high $\beta$, and low $\epsilon_\mathrm{ff}$ end. We are currently working to identify clouds as far into this regime as possible. Combining newly available high resolution and sensitivity sub-mm interferometry data (Longmore et al in prep; Battersby et al submitted) of such clouds with detailed hydrodynamic simulations of gas clouds on the known orbit in the Galactic Centre environment (Kruijssen et al. in prep.), as well as accounting for a more detailed propagation of the observational uncertainties on the dimensionless ratios (taking into account the covariance of uncertainties), we hope to unambiguously distinguish between the competing theories. We point out that the falsification of these star formation theories is currently obstructed by the lack of consensus on the values of their free parameters. 

%{\bf In future, we hope to also expand the comparison presented here to other available star formation models (e.g. \citealp{padoan_2012, zamora-aviles_2012, zamora-aviles_2014}).}

\vspace{5mm}

To summarise, we suggest that the total {\it global (hundred parsec) scale} star formation rate for the Galactic Centre appears to be overpredicted by the star formation models \citep{longmore_2013}, as the majority of the gas is unbound (super-virial), despite it being very dense \citep{kruijssen_2014a}. When investigating {\it local (parsec) scales} within gravitationally bound clouds, we find that several of the models accurately predict the star formation rate. However, a consensus on the free parameters of these models is required before reliable comparison to observations are possible.

\begin{table*}
\centering
\caption{The properties of the sources within the $0.18<l<$0.76$^{\circ}$, $-0.12<b<$0.13$^{\circ}$ region. These do not depend on any relationship between these regions, other than that they all reside at the same distance. The columns show the column density limits used to define the sources, the masses and the radii, bolometric luminosities, maximum embedded object mass, total embedded stellar mass, the free-fall time and the star formation efficiency$^{1}$. Shown in parentheses are the embedded stellar masses, and the resulting cloud properties, determined from VLA, and WMAP observations (see corresponding footnotes for references).}
\begin{tabular}{c c c c c c c c c }
\hline

Source & $N_{H_2}$ boundary & M$_\mathrm{gas}$$^{1}$ & R & L$_\mathrm{bol}$$^{1}$  & M$_\mathrm{*, max}$ & M$_\mathrm{*, tot}$$^{1}$ & $t_\mathrm{ff}$$^{1}$ & SFE$^{1}$ \\
 & (cm$^{-2}$)& ($10^4$\,\sol)& (pc)  & ($10^5$\,L$_\odot$) & (\sol) & (\sol) & (Myr) & (per cent)\\
\hline 

``Brick" & \colhtwocool\,=\,$\,8\,\times10^{22}$ & 11 & 3.1 & 9.1 & 80 &  $<$2.1\,$\times10^{3}$ (8.8\,$\times10^{2}$)$^a$ & 0.27 & 2 \\
``b" & \colhtwocool\,=\,$\,5\,\times10^{22}$ & 1.5 & 1.8 & 2.2 & 40 & $<$9.3\,$\times10^{2}$ & 0.31 & 6 \\
``c" & \colhtwocool\,=\,$\,10\,\times10^{22}$ & 0.51 & 0.8 & 0.59 & 20 & $<$4.2\,$\times10^{2}$ & 0.16 & 8 \\
``d" & \colhtwocool\,=\,$\,10\,\times10^{22}$ & 4 & 2.0 & 2.4 & 40 & $<$9.8\,$\times10^{2}$ & 0.23 & 2 \\
``e" & \colhtwocool\,=\,$\,24\,\times10^{22}$ & 4.8 & 1.5 & 1.8 & 40 & $<$8.1\,$\times10^{2}$ & 0.13 & 2 \\
``f" & \colhtwocool\,=\,$\,24\,\times10^{22}$ & 0.9 & 0.7 & 0.36 & 20 & $<$3.2\,$\times10^{2}$ & 0.10 & 3 \\
Sgr B2 &\colhtwocool\,=\,$\,75\,\times10^{22}$ & 65 & 2.7 & 23 & 120 & $>$3.6\,$\times10^{3}$ (3.3\,$\times10^{4}$)$^b$ & 0.09 (0.09) & 1 (5) \\
G0.6 & \colhtwowarm\,=\,$\,2.6\,\times10^{17}$ & 4.6 & 2.8 & 15 & 100 & 2.8\,$\times10^{3}$ (3.3\,$\times10^{3}$)$^c$ & 0.35 (0.35) & 6 (7) \\
Sgr B1 & \colhtwowarm\,=\,$\,2.6\,\times10^{17}$ & 8.7 & 5.8 & 66 & 180 & 6.0\,$\times10^{3}$ (7.2\,$\times10^{3}$, 8.0\,$\times10^{3}$)$^{c, d}$  & 0.77 (0.77) & 6 (8) \\
G0.3 & \colhtwowarm\,=\,$\,1.9\,\times10^{17}$ & 9.3 & 6.5 & 69 & 180 & 6.2\,$\times10^{3}$ & 0.86 & 6 \\

\hline

\end{tabular}
{\vspace{0.3cm}}
\begin{minipage}{\textwidth}
\vspace{1mm}
$^a$ \citet{kauffmann_2016a} \\
$^b$ Mass determined from high-resolution VLA observations \citet{schmiedeke_2016}. \\
$^c$ Mass determined from medium-resolution VLA observations \citet{mehringer_1992}. \\ 
$^d$ Mass determined from WMAP observations \citet{lee_2012}. \\
$^{1}$ These values represent the instantaneous source properties: M$_\mathrm{gas}$, M$_\mathrm{*, tot}$, $t_\mathrm{ff}$, SFE (see section\,\ref{sec:SFE}).
\end{minipage}
\label{masses_table}
\end{table*}

\begin{table*}
\centering
\caption{The source properties based on the assumptions adopted in section\,\ref{sub:SFE_triggered}. The first assumption is that the``dust-ridge'' clouds are representative of the early evolutionary stages of the star-forming clouds, Sgr B2, G0.6, Sgr B1 and G0.3. The second assumption is that the clouds are on an elliptical orbit around the CMZ, which tidally triggers star formation at the point of pericentre passage. Tabulated in the first column are the times since pericentre passage (i.e. triggering of star formation) as defined from the orbital model of \citep{kruijssen_2015}, which have been used to determine the star formation rates and star formation efficiencies per free-fall time. Shown in parentheses are the properties determined when using the embedded stellar masses calculated from VLA and WMAP observations (see Table\,\ref{masses_table}). }
\begin{tabular}{c c c c c c c c c}
\hline
Source & $t _\mathrm{p,last}$ & SFR & $\epsilon_\mathrm{ff}$ \\
 & (Myr) & (\solyr) & (per cent) \\
\hline 

``Brick"  & 0.3 & 0.007 & 2 &  \\
``b"  & 0.4 & 0.002 & 5 & \\
``c" & 0.4 & 0.001 & 3 & \\
``d" & 0.5 & 0.002 & 1 & \\
``e" & 0.5 & 0.002 & 0.5 &\\
``f'' & 0.5 & 0.001 & 1 &\\
Sgr B2 & 0.7 & 0.005 (0.045) & 0.1 (1)  \\
G0.6  & 1.4 & 0.002 (0.002) & 1 (2) \\
Sgr B1 & 1.6 & 0.004 (0.005) & 3 (4)  \\
G0.3 & 1.8 & 0.004 & 3 \\
\hline
\end{tabular}
\label{SFR_table}
\end{table*}

\section*{ACKNOWLEDGEMENTS}

We would like to thank the anonymous referee for their constructive comments. We greatly appreciate discussions with Gilles Chabrier, Paolo Padoan, Patrick Hennebelle and Christoph Federrath, who provided detailed feedback on the comparison to the volumetric models. Furthermore, we thank Christoph Federrath for providing the probability distribution function of turbulence driving parameter from \citet{federrath_2016}. ATB would like to acknowledge the funding provided by Liverpool John Moores University and the Max Planck Institute for Extraterrestrial Physics. JMDK gratefully acknowledges funding in the form of an Emmy Noether Research Group from the Deutsche Forschungsgemeinschaft (DFG), grant number KR4801/1-1. This research has made use of NASA's Astrophysics Data System.

\bibliographystyle{mnras}
\bibliography{references}
\appendix

\section{Bolometric luminosity as a function of galactic longitude.}\label{Appendix A}

\begin{figure*}
\centering
\includegraphics[trim = 10mm 0mm 10mm 0mm, clip,angle=0,width=1\textwidth]{\dir 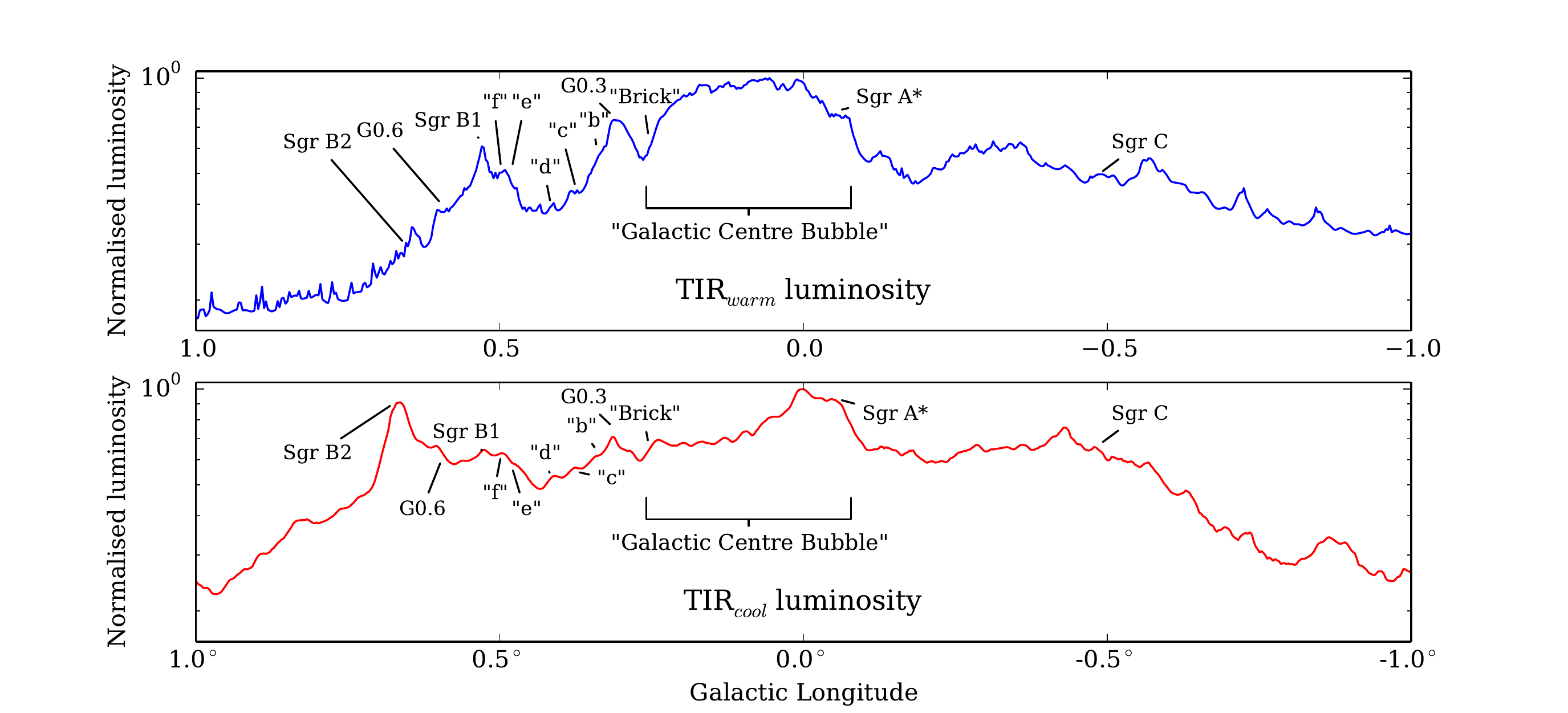}
\caption{Plots the warm (upper; blue) and cool (lower; red) component luminosities as a function of galactic longitude. Positions of sources interest are shown, see Figure\,\ref{sfr_maps}.} 
\label{lum_b}
\end{figure*}

In section\,\ref{method} we calculated the bolometric luminosity towards the Galactic Centre, by fitting a two (cool and warm) component modified black-body to the spectral energy distribution at each position. Figure\,\ref{lum_b} shows how the normalised luminosities for these components vary as a function of galactic longitude for the region $|l|$$<$1$^{\circ}$, $|b|<$0.5$^{\circ}$. The total luminosities across this region is L(TIR$_{warm})\,=\,2.8\,\times\,10^{8}$\,\Lsol\ and L(TIR$_{cool})\,=\,2.9\,\times\,10^{8}$\,\Lsol. We find that 50 and 52 per cent of these luminosities, respectively, are found at positive longitudes ($0<l<1^{\circ}$). 

Shown in Figure\,\ref{lum_b} are several of the sources of interest within the region. For the investigation of the individual star formation rates, we measured the properties for sources within the $0.18<l<$0.76$^{\circ}$, $-0.12<b<$0.13$^{\circ}$ (see Figure\,\ref{rgb}). Table\,\ref{lums} displays the infrared luminosities and the fractions of the total luminosity across the $|l|$$<$1$^{\circ}$, $|b|<$0.5$^{\circ}$ region, for the sources which have not be previously noted.

\begin{table}
\centering
\caption{Luminosities, and the fractions of the total luminosity across the $|l|$$<$1$^{\circ}$, $|b|<$0.5$^{\circ}$ region ($5.7\,\times\,10^{8}$\,\Lsol), for the sources which have not be previously noted (see Figure\,\ref{sfr_maps}).}
\begin{tabular}{c c c}
\hline
Source & L(TIR$_{tot}$) & Fraction of total\\
& (L$_\odot$) & (per cent)\\
\hline
``Galactic Centre Bubble'' & $8.9\,\times\,10^{7}$ & 20\\
Sgr A$^*$ & $3.6\,\times\,10^{6}$ & 0.6\\
Sgr C & $3.1\,\times\,10^{6}$ & 0.5\\
Arches cluster & $2.7\,\times\,10^{5}$ & 0.05 \\
Quintuplet cluster & $2.5\,\times\,10^{5}$ & 0.04\\

\hline
\end{tabular}
\label{lums}
\end{table}

\section{Mass-Bolometric luminosity function.}\label{Appendix B}

\begin{figure}
\centering
\includegraphics[trim = 0mm 0mm 0mm 0mm, clip,angle=0,width=1\columnwidth]{\dir 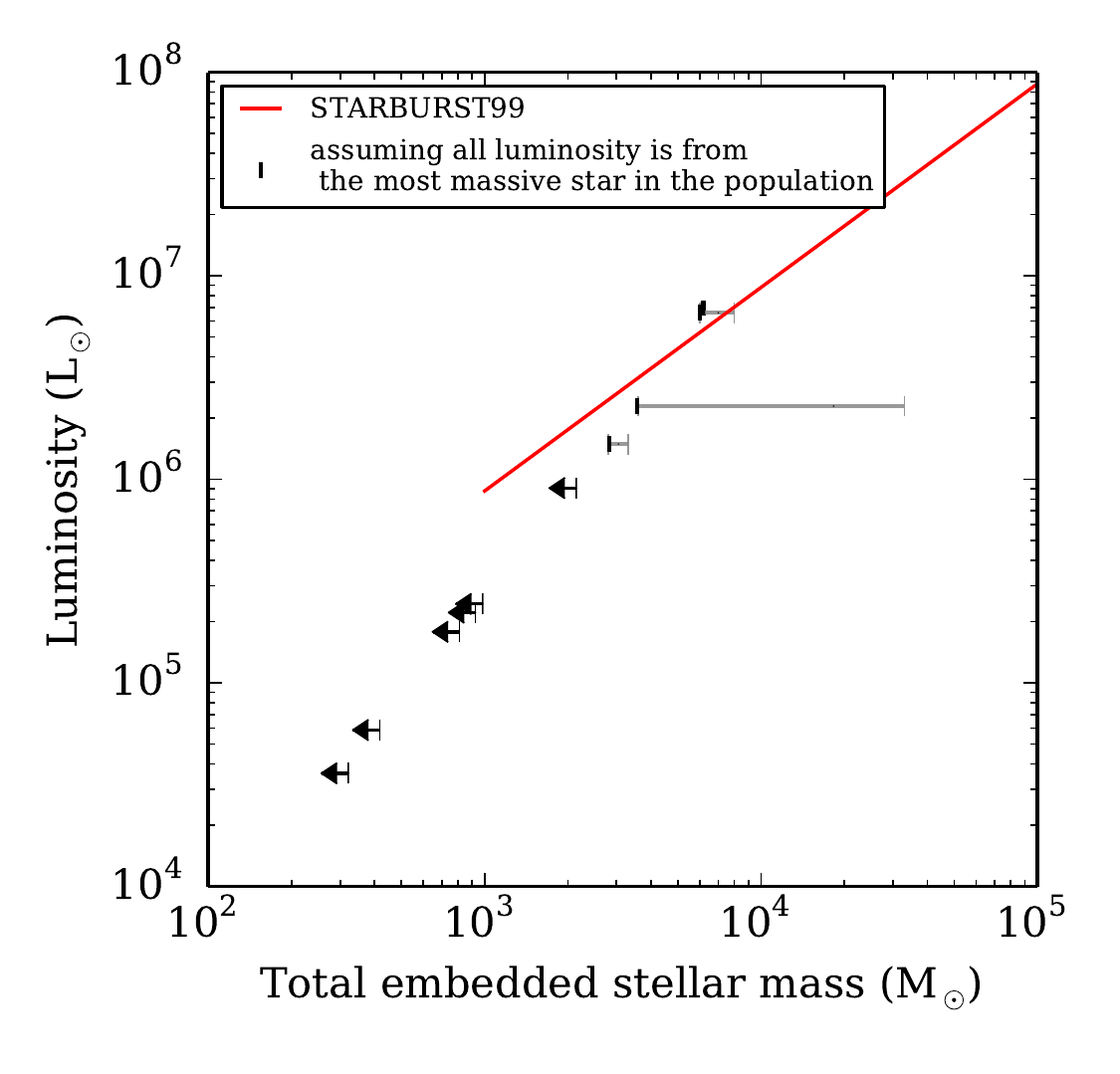}
\caption{Plot displaying the luminosity as a function of mass for the sources in this work when assuming all the luminosity originates from only the most massive embedded star (black lines). The grey error bars indicate the ranges of embedded masses determined from alternative measurements (see section\,\ref{lit_masses}). Also shown is the results of the {\sc starburst99} stellar population model (red line). From this, we conclude that the contribution from lower mass stars to the total bolometric luminosity is insignificant for the larger star-forming sources of our sample (Sgr B2, G0.6, Sgr B1 and G0.3). However, their contribution to the luminosity of the more quiescent clouds (``dust-ridge'' clouds), may result in over estimation of their embedded stellar mass by factors of a few.}
\label{mass_lum}
\end{figure}

In section\,\ref{embedded} we determined the mass of the sources within the $0.18<l<$0.76$^{\circ}$, $-0.12<b<$0.13$^{\circ}$ region, under the assumption that all the bolometric luminosity originates from one massive star. However, as mentioned in Section\,\ref{IMF}, there may be a significant contribution from the lower mass stars in the population. To investigate this we use the synthetic stellar population model {\sc starburst99}\footnote{ \url{http://www.stsci.edu/science/starburst99/docs/default.htm}}. We model an instantaneous star formation burst populating an initial mass function with powers $\alpha$\,=\,0.3 for 0.001\,$<$\,$m/$\,\sol\,$<$\,0.08, $\alpha$\,=\,1.3 for 0.08\,$<$\,$m/$\,\sol\,$<$\,0.5, and $\alpha$\,=\,2.3 for $m$\,$>$\,0.5\,\sol\ \citep{kroupa_2001}, with total cluster masses of 1000 to 100,000\,\sol. All other parameters are left as defaults (e.g. solar metallically). Figure\,\ref{mass_lum} shows the luminosity of the sources measured in this work and those determined from the {\sc starburst99} stellar population model, against their embedded stellar masses. From this, we conclude that the contribution from lower mass stars to the total bolometric luminosity will be minimal for the sources with embedded stellar masses $>\,5000-10000$\,\sol (Sgr B2, G0.6, Sgr B1 and G0.3). However, the contribution from the lower mass stars to the luminosity of the more quiescent clouds (``dust-ridge'' clouds), may result in over estimation of their embedded stellar mass by factors of around $\sim\,2-3$. It is not possible, however, to accurately determine this overestimation for low number statistics.

\section{Changing the parameters in the calculation of $\epsilon_\mathrm{ff}$.}\label{Appendix C}

\begin{figure*}
\centering
\includegraphics[trim = 0mm 7mm 0mm 3mm, clip,angle=0,width=0.86\textwidth]{\dir 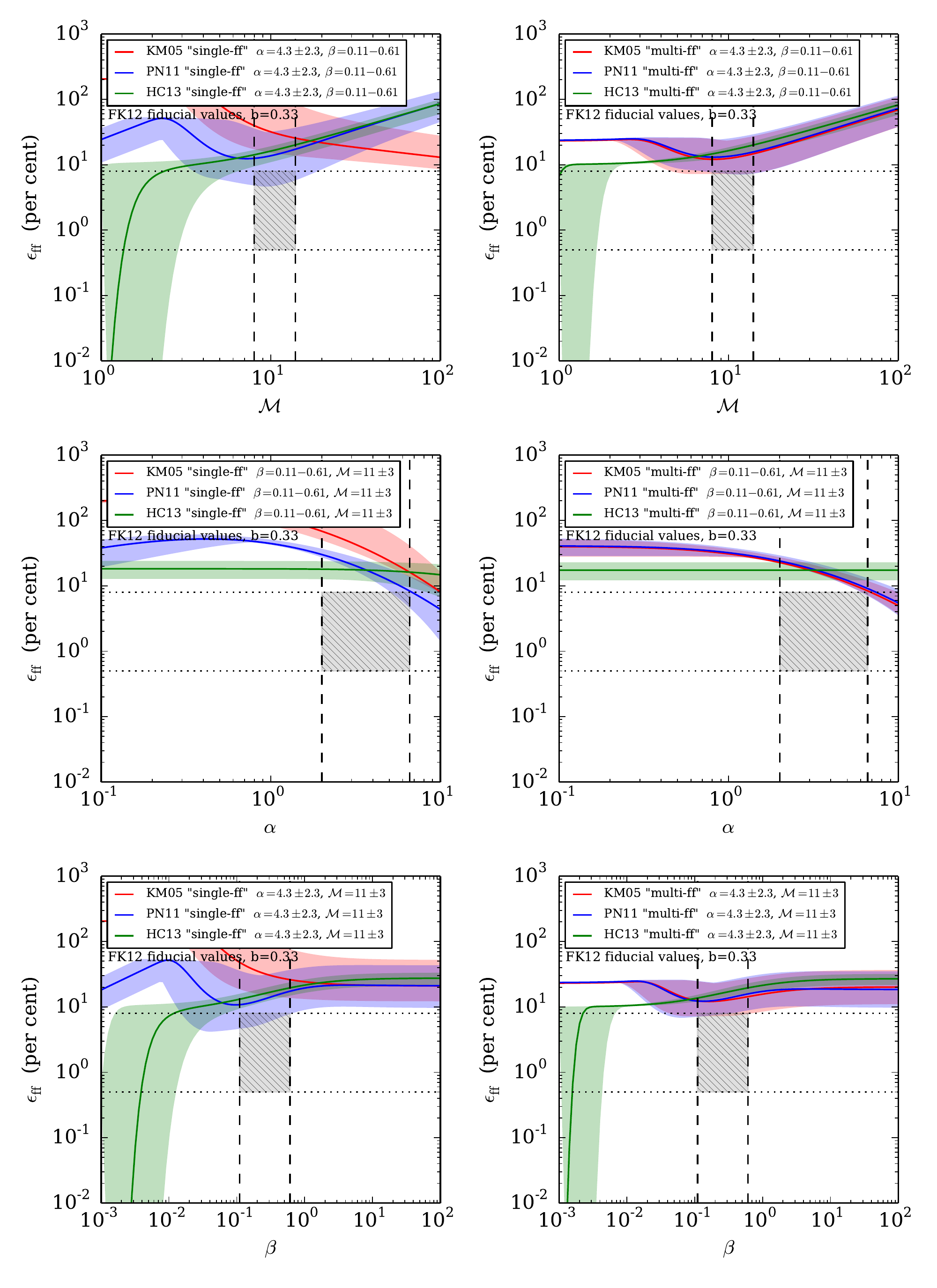}
\caption{This figure is identical to Figure\,\ref{sfr_ff_vs_beta}, however here we use the fiducial values determined by \citet{federrath_2012}. 
For the single-free-fall models these are of $\phi_x\approx0.17$ and $\phi_t$\,=\,0.2 for \citetalias{krumholz_2005}, $\theta\approx0.7$ and $\phi_t$\,=\,0.7 for \citetalias{padoan_2011}, and $y_\mathrm{cut}\approx4.5$ and $\phi_t$\,=\,4.8 for \citetalias{hennebelle_2013}.
 For the multi-free-fall models these are of $\phi_x\approx0.17$ and $\phi_t$\,=\,2.2 for \citetalias{krumholz_2005}, $\theta\approx1.0$ and $\phi_t$\,=\,2.1 for \citetalias{padoan_2011}, and $y_\mathrm{cut}\approx5.9$ and $\phi_t$\,=\,5.0 for \citetalias{hennebelle_2013}.
} 
\label{sfe_ff_models-vol-FK12values-b033}
\end{figure*}

\begin{figure*}
\centering
\includegraphics[trim = 0mm 7mm 0mm 3mm, clip,angle=0,width=0.86\textwidth]{\dir 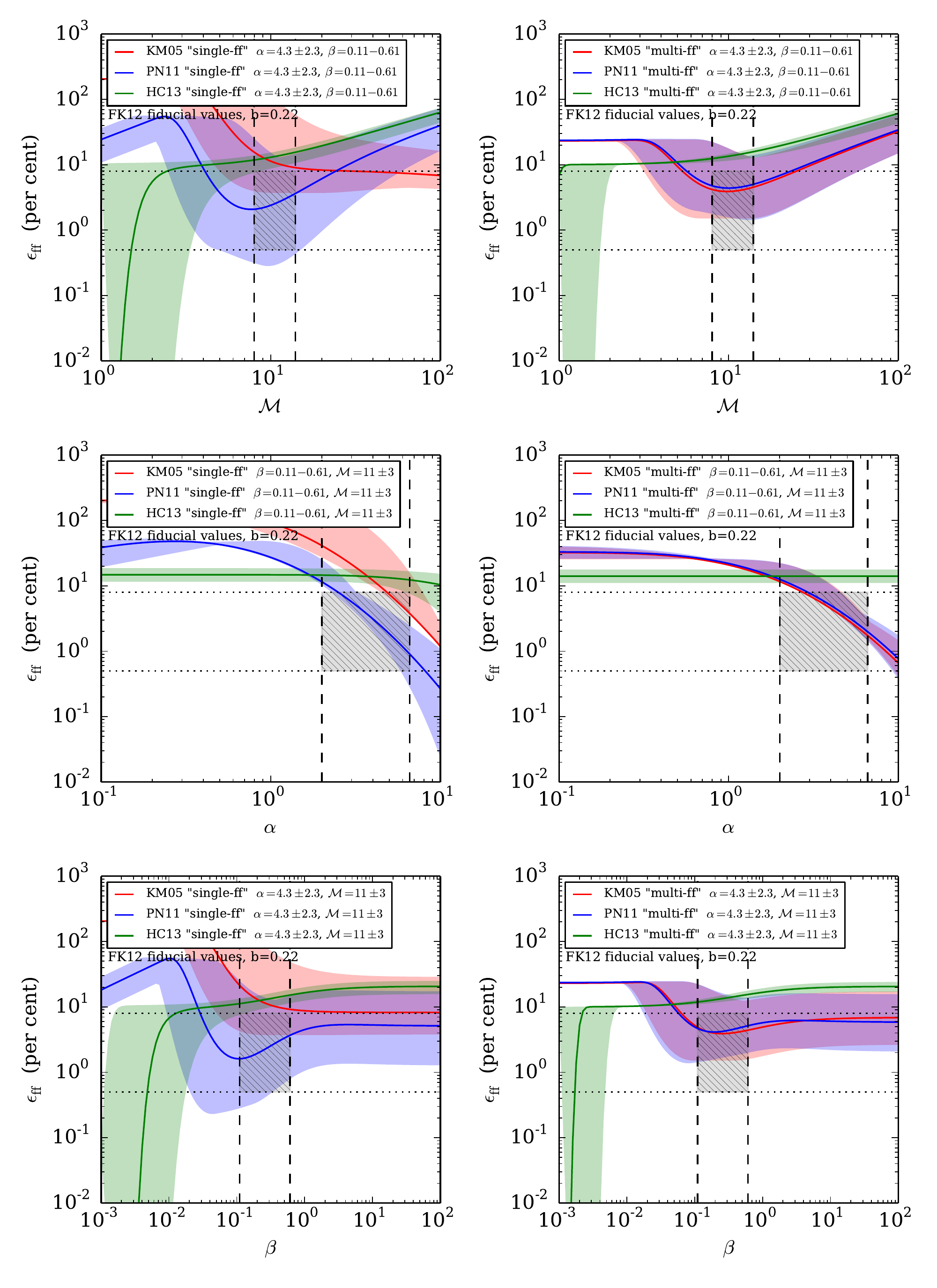}
\caption{This figure is identical to Figure\,\ref{sfr_ff_vs_beta}, however here we use the fiducial values determined by \citet{federrath_2012} and $b = 0.22$. For the single-free-fall models these are of $\phi_x\approx0.17$ and $\phi_t$\,=\,0.2 for \citetalias{krumholz_2005}, $\theta\approx0.7$ and $\phi_t$\,=\,0.7 for \citetalias{padoan_2011}, and $y_\mathrm{cut}\approx4.5$ and $\phi_t$\,=\,4.8 for \citetalias{hennebelle_2013}.
 For the multi-free-fall models these are of $\phi_x\approx0.17$ and $\phi_t$\,=\,2.2 for \citetalias{krumholz_2005}, $\theta\approx1.0$ and $\phi_t$\,=\,2.1 for \citetalias{padoan_2011}, and $y_\mathrm{cut}\approx5.9$ and $\phi_t$\,=\,5.0 for \citetalias{hennebelle_2013}.
} 
\label{sfe_ff_models-vol-FK12values-b022}
\end{figure*}

Figures\,\ref{sfe_ff_models-vol-FK12values-b033} and \ref{sfe_ff_models-vol-FK12values-b022} show the star formation efficiency per free-fall time, as predicted from the single-free-fall (left panels) and multi-free-fall (right panels) models, as a function of the Mach number (upper panels), the virial parameter (middle panels), and the magnetic field strength (lower panels). Figure\,\ref{sfe_ff_models-vol-FK12values-b033} shows the results when adopting the fiducial parameters from \citet{federrath_2012}, and the same turbulent driving parameter used for Figure\,\ref{sfr_ff_vs_beta} ($b = 0.33$), which represents a scenario of purely solenoidal turbulence driving. Figure\,\ref{sfe_ff_models-vol-FK12values-b022} shows the results when adopting the identical parameter set that \citet{federrath_2016} used to predict the $\epsilon_\mathrm{ff}$ within the ``Brick''. These are the best-fitting values of the fiducial parameters from \citet{federrath_2012} (see Figure caption), and a turbulent driving parameter of $b = 0.22$. Table\,\ref{fiducial values} summarises the fiducial values used in the models to create these Figures, and those used in Figure\,\ref{sfr_ff_vs_beta}.

\begin{figure*}
\centering
\includegraphics[trim = 0mm 0mm 0mm 3mm, clip,angle=0,width=0.86\textwidth]{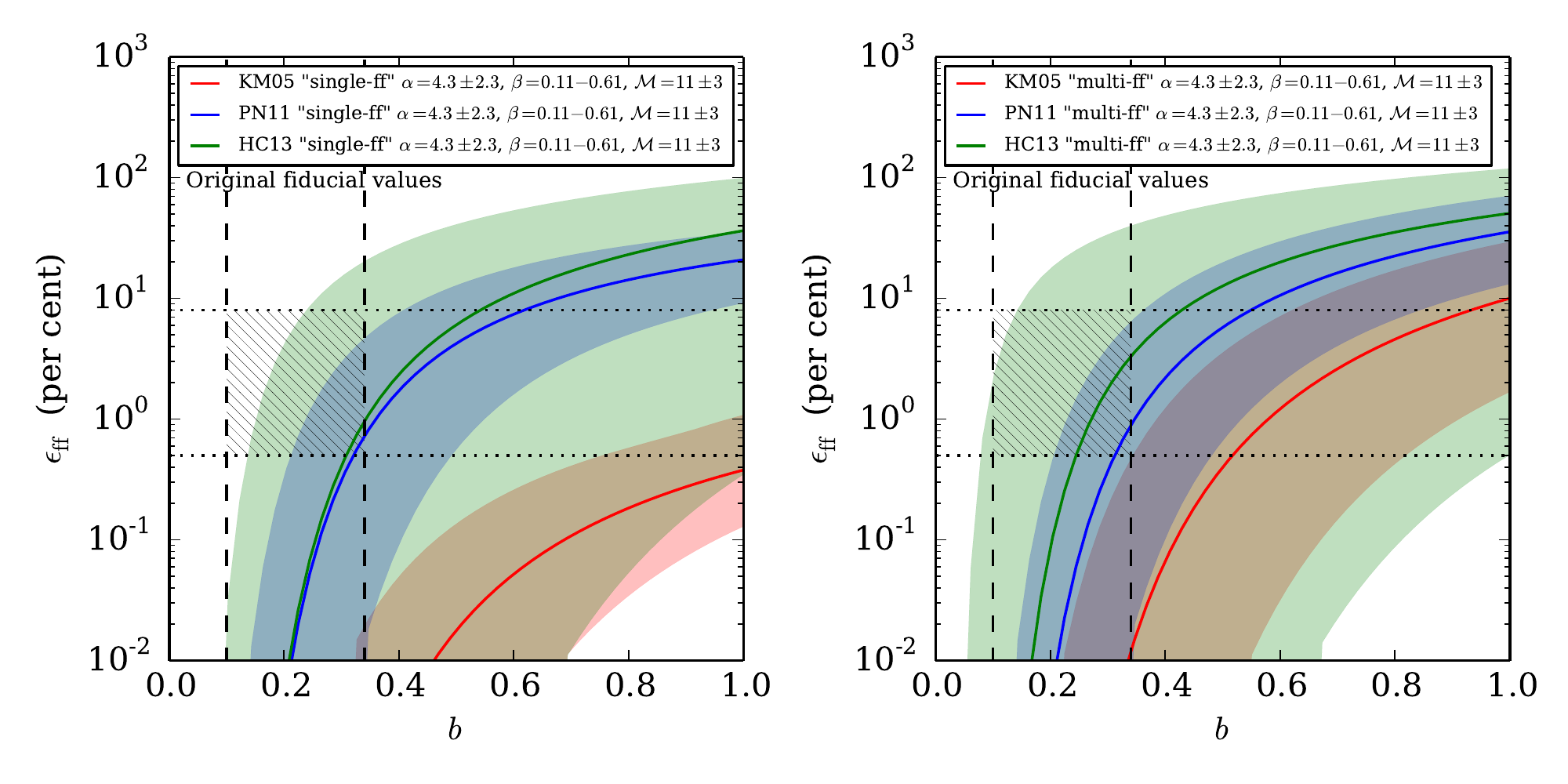}
\includegraphics[trim = 0mm 0mm 0mm 3mm, clip,angle=0,width=0.86\textwidth]{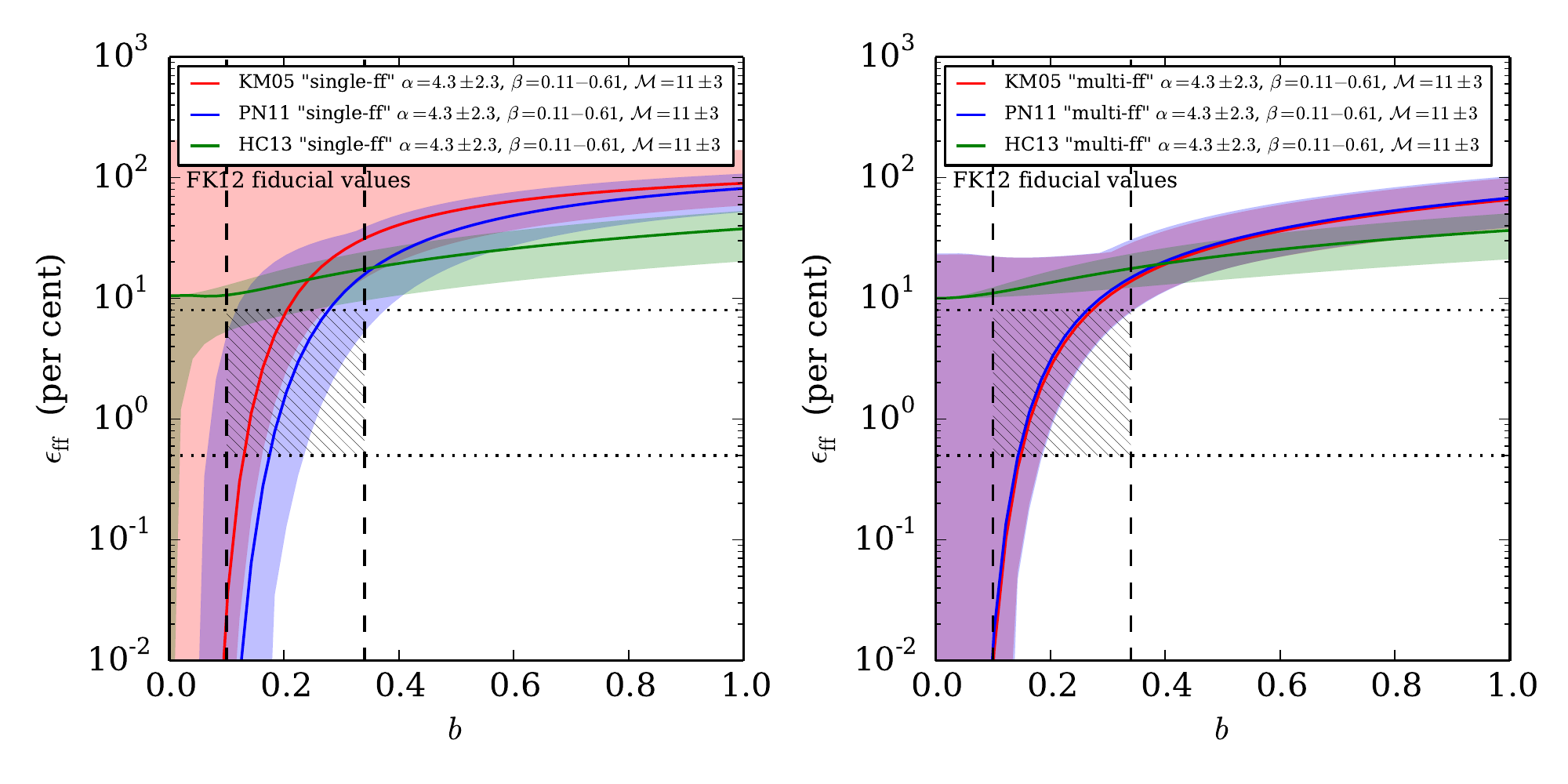}
\caption{Plots of the star formation efficiency per free-fall time, $\epsilon_\mathrm{ff}$, as predicted from the single-free-fall (left panels) and multi-free-fall (right panels) models, as a function of turbulence driving parameter. In the upper panels we assume the fiducial values found by \citetalias{krumholz_2005}, \citetalias{padoan_2011} and \citetalias{hennebelle_2013}, as in Figure\,\ref{sfr_ff_vs_beta}. In the lower panels we assume fiducial values determined by \citet{federrath_2012}, which are used in Figures\,\ref{sfe_ff_models-vol-FK12values-b033} and \ref{sfe_ff_models-vol-FK12values-b022}. The fiducial values are summarised in Table\,\ref{fiducial values}. The coloured lines represent the model predictions. The shaded coloured regions represent the upper and lower limits within the range of our adopted initial conditions (as shown in the legend of each plot). The vertical dashed lines show the result of varying the variable on the x-axis by the range assumed to represent the turbulent driving parameter within the ``Brick'' of $b = 0.22\,\pm\,0.12$ \citep{federrath_2016}. The horizontal dotted regions represent the range of $\epsilon_\mathrm{ff}$ for the star-forming sources within the $0.18<l<$0.76$^{\circ}$, $-0.12<b<$0.13$^{\circ}$ region (determined from both infrared and VLA embedded stellar mass estimates, see Table\,\ref{SFR_table})\textsuperscript{\ref{footnote1}}, accounting for the approximate factor of two uncertainty in $\epsilon_\mathrm{ff}$ (i.e. $\epsilon_\mathrm{ff}$ = 0.5 -- 8 per cent; see section\,\ref{Uncertainties}).}
\label{sfe_ff_models-vol-b}
\end{figure*}

Figure\,\ref{sfe_ff_models-vol-b} shows the star formation efficiency per free-fall time, as predicted from the single-free-fall (left panels) and multi-free-fall (right panels) models, as a function of the turbulence driving parameter. Here we plot the results assuming the fiducial values determined by \citetalias{krumholz_2005}, \citetalias{padoan_2011} and \citetalias{hennebelle_2013}, and those determined by \citet{federrath_2012}.

\begin{table*}
\centering
\caption{Summary of parameters used for in the volumetric models shown in Figures\,\ref{sfr_ff_vs_beta}, \ref{sfe_ff_models-vol-FK12values-b033}, and \ref{sfe_ff_models-vol-FK12values-b022}.}
\begin{tabular}{c c c c c c c c c c c c c c c c}
\hline
Figure & \multicolumn{14}{c}{Fiducial values} \\
 & \multicolumn{2}{c}{All models} & \multicolumn{2}{c}{\citetalias{krumholz_2005}} & \multicolumn{2}{c}{\citetalias{padoan_2011}} & \multicolumn{2}{c}{\citetalias{hennebelle_2013}} & \multicolumn{2}{c}{\citetalias{krumholz_2005}} & \multicolumn{2}{c}{\citetalias{padoan_2011}} & \multicolumn{2}{c}{\citetalias{hennebelle_2013}} \\

 & & & \multicolumn{2}{c}{single-free-fall} & \multicolumn{2}{c}{single-free-fall} & \multicolumn{2}{c}{single-free-fall} & \multicolumn{2}{c}{multi-free-fall} & \multicolumn{2}{c}{multi-free-fall} & \multicolumn{2}{c}{multi-free-fall} \\
 
 \hline
 &$b$ & $\epsilon_{core}$ & $\phi_t$ & $\phi_x$ & $\phi_t$ & $\theta$ & $\phi_t$ & $y_\mathrm{cut}$ & $\phi_t$ & $\phi_x$ & $\phi_t$ & $\theta$ & $\phi_t$ & $y_\mathrm{cut}$\\
\ref{sfr_ff_vs_beta} & 0.33 & 0.5 & 1.19 & 1.12  & 1.19 & 0.35 & 1.19 & 0.1 & 1.19 & 1.12  & 1.19 & 0.35 & 1.19 & 0.1\\
\ref{sfe_ff_models-vol-FK12values-b033} & 0.33 & 0.5 & 0.2 & 0.17  & 0.7 & 0.7 & 4.8 & 4.5 & 2.2 & 0.17 & 2.1 & 1.0 & 5.0 & 5.9\\
\ref{sfe_ff_models-vol-FK12values-b022} & 0.22 & 0.5 & 0.2 & 0.17  & 0.7 & 0.7 & 4.8 & 4.5 & 2.2 & 0.17 & 2.1 & 1.0 & 5.0 & 5.9\\
\hline
\end{tabular}
\label{fiducial values}
\end{table*}

\label{lastpage}
\end{document}